\def\ifhtx{\iffalse}    
\ifhtx
\else
\documentclass[12pt,notitlepage,onecolumn]{ivoa}
\fi

\newif\iftth
\iftth
\definecolor{ivoacolor}{rgb}{0.0,0.318,0.612}
\newcommand{\tthmaketitle}{\par
  \begingroup
    \renewcommand\thefootnote{\@fnsymbol\c@footnote}%
    \def\@makefnmark{\rlap{\@textsuperscript{\normalfont\@thefnmark}}}%
    \long\def\@makefntext##1{\parindent 1em\noindent
            \hb@xt@1.8em{%
                \hss\@textsuperscript{\normalfont\@thefnmark}}##1}%
    \@maketitle
    \thispagestyle{plain}\@thanks
  \endgroup
  \setcounter{footnote}{0}%
  \global\let\thanks\relax
  \global\let\maketitle\relax
  \global\let\@maketitle\relax
  \global\let\@thanks\@empty
  \global\let\@editor\@empty
  \global\let\@author\@empty
  \global\let\@date\@empty
  \global\let\@version\@empty
  \global\let\@title\@empty
  \global\let\title\relax
  \global\let\editor\relax
  \global\let\author\relax
  \global\let\date\relax
  \global\let\version\relax
  \global\let\and\relax
}
\newcommand{\ivoatype}[1]{\def\@ivoatype{#1}}
\newcommand{\ivoagroup}[1]{\def\@ivoagroup{#1}}
\newcommand{\version}[1]{\def\@version{#1}}
\newcommand{\editor}[1]{\def\@editor{#1}}
\newcommand{\author}[1]{\def\@author{#1}}
\newcommand{\date}[1]{\def\@date{#1}}
\newcommand{\title}[1]{\def\@title{#1}}
\newcommand{\urlthisversion}[1]{\def\@urlthisversion{#1}}
\newcommand{\urllastversion}[1]{\def\@urllastversion{#1}}
\newcommand{\previousversion}[1]{\def\@previousversion{#1}}
\ivoatype{}
\version{}
\editor{}
\urlthisversion{}
\urllastversion{}
\previousversion{}
\def\@maketitle{%
  \vskip 1.0em%
  \let \footnote \thanks%
    {\Large\color{ivoacolor}\sffamily\bfseries Version \@version}%
    \vskip 2.0em%
    {\Large\color{ivoacolor}\sffamily\bfseries \@ivoatype{ }\@date}%
    \vskip 2.0em%
    {\large
    \begin{description}
    \item[This version:] {\normalsize\@urlthisversion}
    \item[Latest version:] {\normalsize\@urllastversion}
    \item[Previous versions:] {\normalsize\@previousversion}
    \item[Working Group:] \@ivoagroup
    \item[Authors:] \@author
    \item[Editors:] \@editor
    \end{description}
    }
}
\else
  
  \def\href#1#2{#2}
\fi

\iftth
  \newcommand{\dbreak}{}
  \newcommand{\dbreakd}{}
\else
  \newcommand{\dbreak}{\mbox{}\protect\\}
  \newcommand{\dbreakd}{\mbox{}\protect\\[-3ex]}
\fi
\newcommand{\examplesize}{\small}

\ifx\pdftexversion\undefined 
  \usepackage[dvips]{graphicx}
  \DeclareGraphicsExtensions{.eps,.ps}
  \usepackage[ps2pdf]{hyperref}

\else                        
  \usepackage[pdftex]{graphicx} 
  \usepackage{epstopdf}         
  \makeatletter
  \g@addto@macro\Gin@extensions{,.ps}
  \@namedef{Gin@rule@.ps}#1{{pdf}{.pdf}{`ps2pdf #1}}
  \makeatother
\fi

\usepackage{color}

\newcommand{\sampversion}{1.3}
\newcommand{\sampdate}{2012-04-11}
\newcommand{\samptype}{Recommendation}

\title{SAMP --- Simple Application Messaging Protocol}

\date{\sampdate}
\version{\sampversion}

\ivoatype{IVOA Recommendation}

\ivoagroup{Applications}

\author{M.~Taylor (m.b.taylor@bristol.ac.uk)\\
        T.~Boch (boch@astro.u-strasbg.fr)\\
        M.~Fitzpatrick (fitz@noao.edu)\\
        A.~Allan (aa@astro.ex.ac.uk)\\
        J.~Fay (jfay@microsoft.com)\\
        L.~Paioro (luigi@lambrate.inaf.it)\\
        J.~Taylor (jontayler@gmail.com)\\
        D.~Tody (dtody@nrao.edu)}
\editor{T.~Boch, M.~Fitzpatrick, M.~Taylor}

\def\SVN$#1: #2 ${\expandafter\def\csname SVN#1\endcsname{#2}}
\SVN$Rev: 1684 $
\urlthisversion{\sampversion: \samptype\ \sampdate\ (Revision \SVNRev)}
\urllastversion{\url{http://www.ivoa.net/Documents/latest/SAMP.html}}
\previousversion{
   1.0:  Working Draft 2008-06-25\\
   1.11: Recommendation 2009-04-21 (Revision 987)\\
   1.2:  Recommendation 2010-12-16 (Revision 1384)}

\begin{document}

\iftth
  \tthmaketitle
\else
  \maketitle 
  \newpage   
\fi

\begin{abstract}
SAMP is a messaging protocol that enables astronomy software tools to 
interoperate and communicate.

IVOA members have recognised that building a monolithic tool 
that attempts to fulfil all 
the requirements of all users is impractical, and it is a better use of our 
limited resources to enable individual tools to work together better. 
One element of this is defining common file formats for the exchange of data 
between different applications. Another important component is a messaging 
system that enables the applications to share data and take advantage of each 
other's functionality.
SAMP supports communication between applications on the desktop
and in web browsers,
and is also intended to form a framework for more general messaging
requirements.
\end{abstract}

\section*{Status of this Document}

This document has been produced by the IVOA Applications Working Group.
It has been reviewed by IVOA Members and other interested parties,
and has been endorsed by the IVOA Executive Committee as an IVOA
Recommendation.
It is a stable document and may be used as reference material or cited as
a normative reference from another document.  IVOA's role in making the
Recommendation is to draw attention to the specification and to promote
its widespread deployment.  This enhances the functionality and
interoperability inside the Astronomical Community.

Comments, questions and discussions relating to this document
may be posted to the mailing list of the SAMP subgroup of the 
Applications Working Group, 
\href{mailto:apps-samp@ivoa.net}{apps-samp@ivoa.net}.
Supporting material and further discussion
may be found at \url{http://www.ivoa.net/samp/}.

Changes since earlier versions may be found in Appendix \ref{sect:changes}.

\tableofcontents

\section{Introduction}

\subsection{Non-Technical Preamble and Position in IVOA Architecture}
\label{sect:nonTechPreamble}
SAMP, the Simple Application Messaging Protocol, is a standard
for allowing software tools to exchange control and data information,
thus facilitating tool interoperability, and so allowing users
to treat separately developed applications as an integrated suite.
An example of an operation that SAMP might facilitate is passing a
source catalogue from one GUI application to another, and 
subsequently allowing sources marked by the user in one of those 
applications to be visible as such in the other.

The protocol has been designed, and implementations developed, within
the context of the International Virtual Observatory Alliance (IVOA),
but the design is not specific either to the Virtual Observatory (VO)
or to Astronomy.  It is used in practice 
for both VO and non-VO work with astronomical tools, and is in  
principle suitable for non-astronomical purposes as well.

\begin{figure}[tbh]
\begin{center}
\includegraphics[scale=0.55]{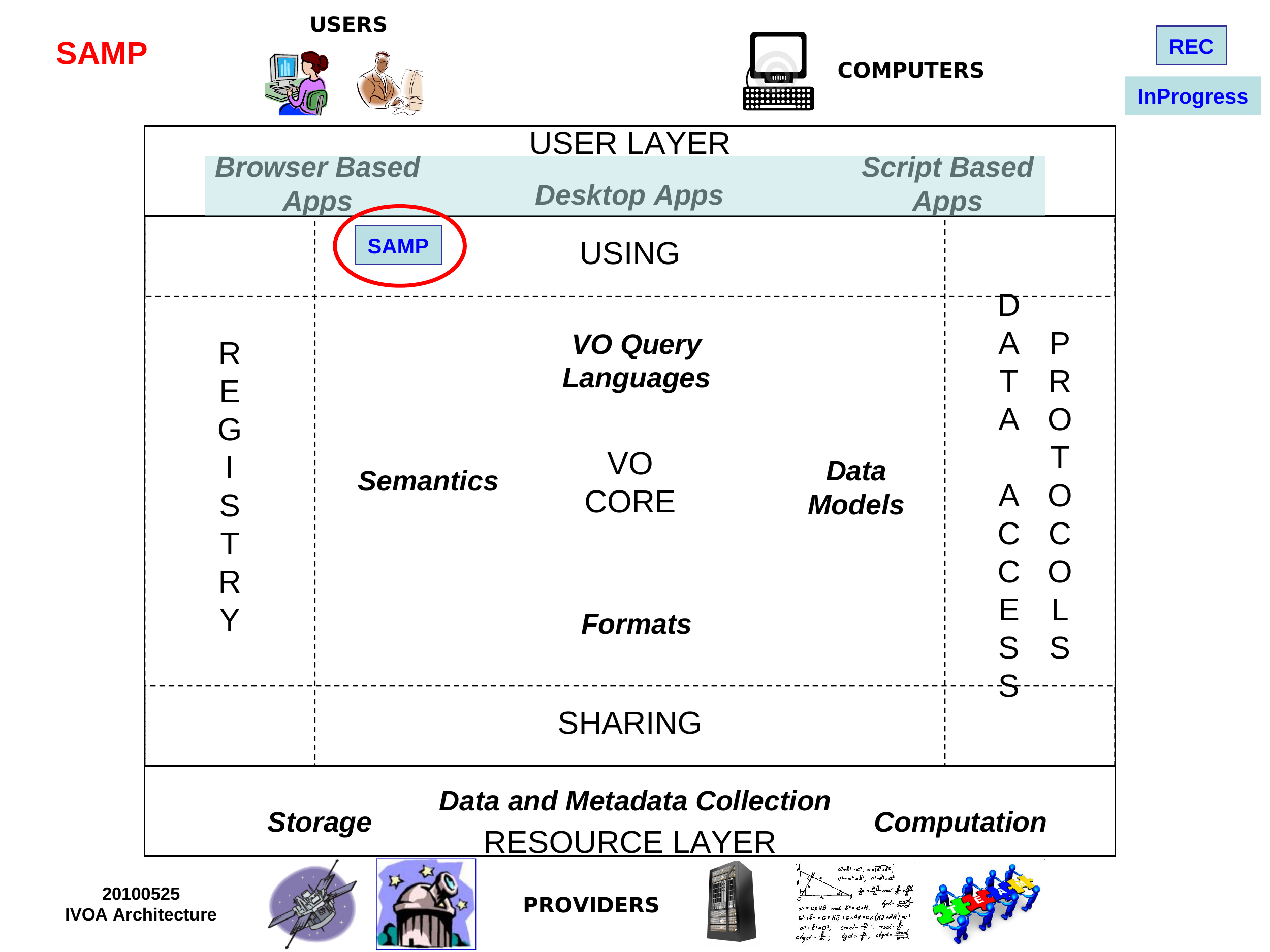}
\caption{IVOA Architecture diagram \cite{architecture}.
The SAMP protocol appears in the ``Using'' region.}
\label{fig:ivoa-archi}
\end{center}
\end{figure}

The SAMP standard itself is neither a dependent, nor a dependency,
of other VO standards, but it provides valuable glue between user-level
applications which perform different VO-related tasks, and hence
contributes to the integration of Virtual Observatory functionality
from a user's point of view.
Figure~\ref{fig:ivoa-archi} illustrates SAMP in the context of the
IVOA Architecture \cite{architecture}.
Most existing tools which operate in the 
User Layer of this architecture provide SAMP interoperability.

The semantics of messages that can be exchanged using SAMP are defined
by contracts known as MTypes (message-types), which are defined
by developer agreement outside of this standard.  The list of MTypes
used for common astronomical and VO purposes can be found near 
\url{http://www.ivoa.net/samp/}; many of these make use of standards from 
elsewhere in the IVOA Architecture, including VOTable, VOResource, 
Simple Spectral Access, UCD and Utype.

\subsection{History}
SAMP, the Simple Application Messaging Protocol,
 is a direct descendent of the PLASTIC protocol, which in turn grew 
--- in the European VOTech framework --- 
from the interoperability work of the Aladin \cite{2000A+AS..143...33B} and 
VisIVO \cite{2007HiA....14..622B} teams. We also note the 
contribution of the team behind the earlier XPA protocol \cite{xpa}. For more information 
on PLASTIC's history and purpose see the IVOA Note {\em PLASTIC --- a protocol for 
desktop application interoperability} \cite{plastic}
and the PLASTIC SourceForge site \cite{plastic-sf}.

SAMP has similar aims to PLASTIC, but 
incorporates lessons learnt from two years of practical experience and ideas 
from partners who were not involved in PLASTIC's initial design. 

Broadly speaking, SAMP is an abstract framework for
loosely-coupled, asynchronous, RPC-like and/or event-based communication,
based on a central service providing multi-directional publish/subscribe
message brokering.
The message semantics are extensible and use structured but weakly-typed data.
These concepts are expanded on below.
It attempts to make as few assumptions as possible about the transport
layer or programming language with which it is used.
It also defines a ``Standard Profile'' which specifies how to 
implement this framework using XML-RPC \cite{xmlrpc} as the transport layer.
The result of combining this Standard Profile with the rest of the SAMP
standard is deliberately similar in design to PLASTIC,
and this has been largely successful in its intention of enabling
PLASTIC applications to be modified to use SAMP instead without great effort.
More recently (version 1.3) an additional ``Web Profile'' has been introduced,
in order to facilitate use of SAMP from web applications.

\subsection{Requirements and Scope}
SAMP aims to be a simple and extensible protocol that is platform- and 
language-neutral. 
The emphasis is on a 
simple protocol with a very shallow learning curve in order to encourage
as many application authors as possible to adopt it.
SAMP is intended to do what you need most of the time. The SAMP authors 
believe that this is the best way to foster innovation and collaboration in
astronomy applications.

It is important to note therefore that SAMP's scope is reasonably modest; it is 
not intended to be the perfect messaging solution for all situations.
In particular SAMP itself has 
no support for transactions, security, or guaranteed message delivery 
or integrity.
However, by layering the SAMP architecture on top of suitable
messaging infrastructures such capabilities could be provided.
These possibilities are not discussed further in this document,
but the intention is to provide an architecture which is sufficiently
open to allow for such things in the future with little change to the
basics.

\subsection{Types of Messaging}
\label{sect:typeOfMsging}

        SAMP is currently targetted at inter-application desktop messaging
with the idea that the basic framework presented here is extensible to meet
future needs, and so it is beyond the scope of this document to outline the
many types of messaging systems in use today (these are covered in detail
in many other documents).  While based on established messaging models,
SAMP is in many ways a hybrid of several basic messaging concepts; the
protocol is however flexible enough that later versions should be able to
interact fairly easily with other messaging systems because of the shared
messaging models.

The messaging concepts used within SAMP include:

\begin{description}
\item[Publish/Subscribe Messaging:]
    A publish/subscribe (pub/sub) messaging system supports an event driven
    model where information producers and consumers participate in message
    passing.  SAMP applications ``publish'' a message, while consumer
    applications ``subscribe'' to messages of interest and consume events.
    The underlying messaging system routes messages from producers 
    to consumers based on
    the message types in which an application has registered an interest.
\item[Point-to-Point Messaging:]
    In point to point messaging systems, messages are routed to an
    individual consumer which maintains a queue of ``incoming'' messages.  In
    a traditional message queue, applications send messages to a specified
    queue and clients retrieve them.  In SAMP, the message system manages
    the delivery and routing of messages, but also permits the concept of a
    directed message meant for delivery to a specific application.  SAMP
    does not, however, guarantee the order of message delivery as with a
    traditional message queue.
\item[Event-based Messaging:]
    Event-based systems are systems in which producers generate events, and in
    which messaging middleware delivers events to consumers based upon a
    previously specified interest.  One typical usage pattern of these systems
    is the publish/subscribe paradigm, however these systems are also widely
    used for integrating loosely coupled application components.  SAMP allows
    for the concept that an ``event'' occurred in the system and that these
    message types may have requirements different from messages where the
    sender is trying to invoke some action in the network of applications.
\item[Synchronous vs. Asynchronous Messaging:]
    As the term is used in this document, a ``synchronous'' message is one
    which blocks the sending application from further processing until a
    reply is received.  However, SAMP messaging is based on ``asynchronous''
    message and response in that the delivery of a message and its
    subsequent response are handled as separate activities by the
    underlying system.  With the exception of the synchronous message
    pattern supported by the system, sending or replying to a message using
    SAMP allows an application to return to other processing while the
    details of the delivery are handled separately.
\end{description}

\subsection{About this Document}
This document contains the following main sections describing the SAMP protocol
and how to use it.
Section \ref{sect:architecture}
covers the requirements, basic concepts and overall architecture of SAMP.
Section \ref{sect:apis} defines abstract (i.e.\ independent of language,
platform and transport protocol) interfaces which clients and hubs 
must offer to participate in SAMP messaging, along with data types
and encoding rules required to use them.
Section \ref{sect:profile}
explains how the abstract API can be mapped to specific network operations
to form an interoperable messaging system, and defines the ``Standard Profile'',
based on XML-RPC, which gives a particular set of such mappings
suitable for general purpose desktop applications.
Section \ref{sect:webprofile} defines the ``Web Profile'',
an alternative mapping suitable for web applications.
Section \ref{sect:mtypes}
describes the use of the MType keys used to denote message semantics,
and outlines an MType vocabulary.

The key words ``MUST'', ``MUST NOT'', ``REQUIRED'', ``SHALL'', ``SHALL NOT'', 
``SHOULD'', ``SHOULD NOT'', ``RECOMMENDED'',  ``MAY'', and ``OPTIONAL'' 
in this document are to be interpreted as described in RFC 2119 \cite{rfc2119}.

\section{Architectural Overview}
\label{sect:architecture}
This section provides a high level view of the SAMP protocol.

\subsection{Nomenclature}
In the text that follows these terms are used: 

\begin{description}
\item[Hub:]
    A broker service for routing SAMP Messages.
\item[Client:]
    An application that talks to a Hub using SAMP. May be a
    Sender, Recipient, or both.
\item[Sender:]
    A Client that sends a SAMP Message to one or more
    Recipients via the Hub.
\item[Recipient:]
    A Client that receives a SAMP Message from
    the Hub. This may have originated from another Client or from the
    Hub itself.
\item[Message:]
    A communication sent from a Sender to a Recipient
    via a SAMP Hub.  Contains an MType and zero or more named parameters.
    May or may not provoke a Response.
\item[Response:]
    A communication which may be returned from a Recipient to a Sender
    in reply to a previous Message.  A Response may contain returned values
    and/or error information.  In the terminology of this document,
    a Response is not itself a Message.  A Response is also known as a
    Reply in this document.
\item[MType:]
    A string defining the semantics of a Message and of its arguments and
    return values (if any).  Every Message contains exactly one MType,
    and a Message is only delivered to Clients subscribed to that MType.
\item[Subscription:]
    A Client is said to be Subscribed to a given MType if it has
    declared to the Hub that it is prepared to receive Messages
    with that MType.
\item[Callable Client:]
    A Client to which the Hub is capable of performing callbacks.
    Clients are not obliged to be Callable, but only Callable Clients
    are able to receive Messages or asynchronous Responses.
\item[Broadcast:]
    To send a SAMP Message to all Subscribed Clients 
    excluding the Sender.
\item[Profile:]
    A set of rules which map the abstract API defined by SAMP to a set of
    I/O operations which may be used by Clients to send and receive
    actual Messages.
\end{description}

\subsection{Messaging Topology}
\begin{figure}
\begin{center}
\includegraphics[scale=0.5]{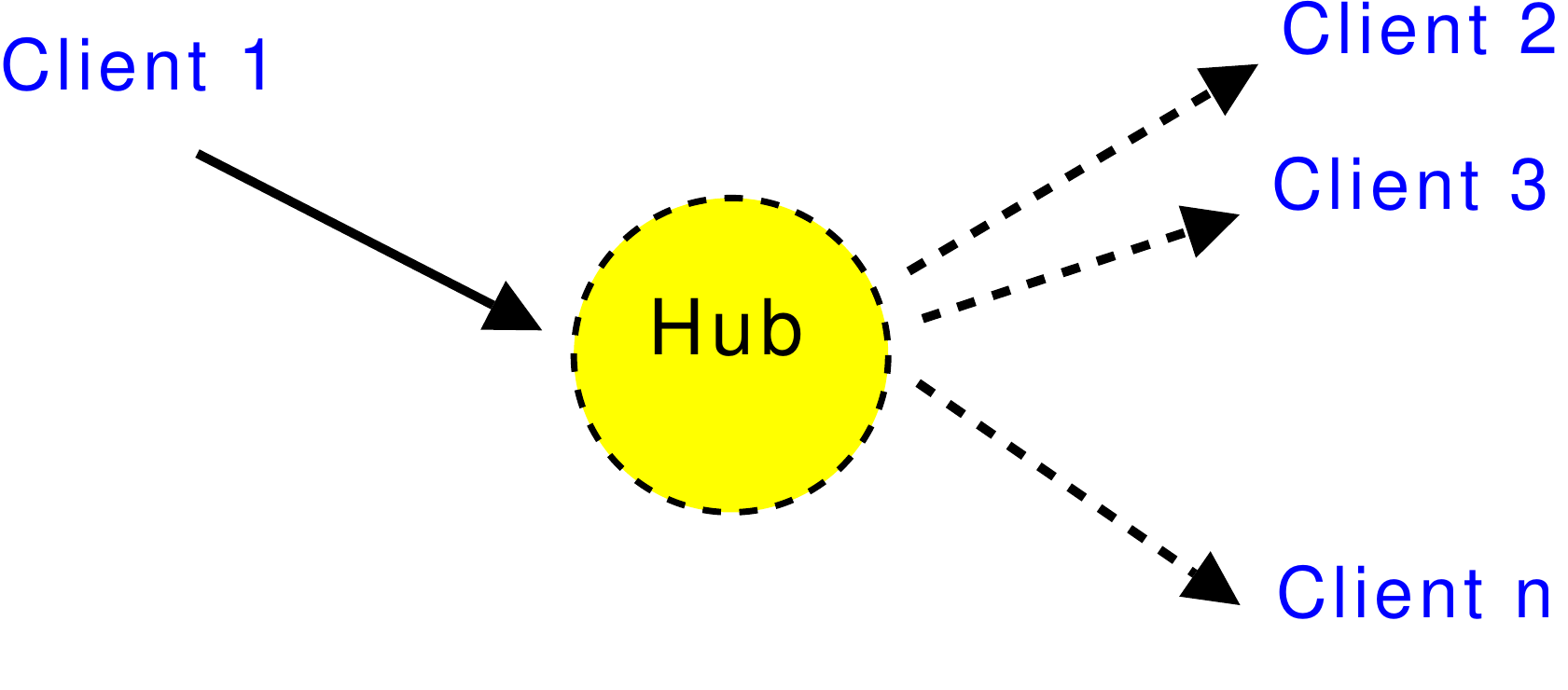}
\caption{The SAMP hub architecture}
\label{fig:samp-archi}
\end{center}
\end{figure}

SAMP has a hub-based architecture (see Figure~\ref{fig:samp-archi}). The hub is a single service used to route all 
messages between clients. This makes application discovery more 
straightforward in that each client only needs to locate the hub, and the services
provided by the hub are intended to simplify the actions of the client. A disadvantage of this 
architecture is that the hub may be a message bottleneck and potential single point of failure. 
The former means that SAMP may not be suitable for extremely high
throughput requirements;
the latter may be mitigated by an appropriate strategy for hub restart if failure is likely.

Note that the hub is defined as a service interface which may have any of
a number of implementations.  It may be an independent application running as
a daemon, an adapter interface layered on top of an existing messaging
infrastructure, or a service provided by an application which is itself
one of the hub's clients.

\subsection{The Lifecycle of a Client}

A SAMP client goes through the following phases: 
\begin{enumerate}
    \item Determine whether a hub is running by using the appropriate hub discovery 
    mechanism.
    \item If so, use the hub discovery mechanism to work out how to communicate
    with the hub.
    \item Register with the hub.
    \item Store metadata such as client name, description and icon in the 
    hub.
    \item Subscribe to a list of MTypes to define messages which may be 
    received.
    \item Interrogate the hub for metadata of other clients.
    \item Send and/or receive messages to/from other clients via the hub.
    \item Unregister with the hub.
\end{enumerate}
Phases 4--7 are all optional and may be repeated in any order. 

By subscribing to the MTypes described in Section \ref{sect:hub-mtypes}
a client may, if it wishes, keep track of the details of other clients' 
registrations, metadata and subscriptions.

\subsection{The Lifecycle of a Hub}
A SAMP hub goes through the following phases: 

\begin{enumerate}
    \item Locate any existing hub by using the appropriate hub discovery mechanism.
    \begin{enumerate}
        \item Check whether the existing hub is alive.
        \item If so, exit.
    \end{enumerate}
    \item If no hub is running, or a hub is found but is not 
    functioning, write/overwrite the hub discovery record and start up.
    \item Await client registrations. When a client makes a legal 
    registration, assign it a public ID,
    and add the client to the table of registered 
    clients under the public ID. Broadcast a message
    announcing the registration of a 
    new client.
    \item When a client stores metadata in the hub, broadcast a message 
    announcing the change
    and make the metadata available.
    \item When a client updates its list of subscribed MTypes, broadcast a 
    message announcing the change and make the subscription information
    available
    \item When the hub receives a message for relaying, pass it on to 
    appropriate recipients which are subscribed to the message's MType.
    Broadcast messages are sent to all 
    subscribed clients except the sender, messages with a specified recipient
    are sent to that recipient if it is subscribed.
    \item Await client unregistrations. When a client unregisters, broadcast 
    a message announcing the unregistration 
    and remove the client from the table of registered clients. 
    \item If the hub is unable to communicate with a client, it may 
    unregister it as described in phase 7.
    \item When the hub is about to shutdown, broadcast a message to 
    all subscribed clients.
    \item Delete the hub discovery record.
\end{enumerate}
Phases 3--8 are responses to events which may occur 
multiple times and in any order.

The MTypes broadcast by the hub to inform clients of changes in its state
are given in Section \ref{sect:hub-mtypes}.

Readers should note that, given this scheme, race conditions may occur.
A client might for instance try to register with a hub which has just shut down,
or attempt to send to a recipient which has already unregistered.
Specific profiles MAY define best-practice rules in order to best manage these
conditions,
but in general clients should be aware that SAMP's lack of guaranteed
message delivery and timing means that unexpected conditions are possible.

\subsection{Message Delivery Patterns}
\label{sect:delivery-outline}

Messages can be sent according to three patterns, differing in 
whether and how a response is returned to the sender:
\begin{enumerate}
    \item Notification
    \item Asynchronous Call/Response
    \item Synchronous Call/Response
\end{enumerate}
The Notification pattern is strictly one-way while in the Call/Response
patterns the recipient returns a response to the sender.

If the sender expects to receive some useful data as a result of the
receiver's processing, or if it wishes to find out whether and
when the processing is completed, it should use one of the Call/Response
variants.  If on the other hand the sender has no interest in what
the recipient does with the message once it has been sent, it
may use the Notification pattern.  Notification, since it involves
no communication back from the recipient to the sender, uses
fewer resources.
Although typically ``event''-type messages will be sent using Notify
and ``request-for-information''-type messages will be sent using
Call/Response, the choice of which delivery pattern to use is
entirely distinct from the content of the message, and is up to
the sender; any message (MType) may be sent using any of the above
patterns.
Apart from the fact of returning or not returning a response,
the recipient SHOULD process messages in exactly the same way 
regardless of which pattern is used.

From the receiver's point of view there are only two cases:
Notification and Asynchronous Call/Response.
However, the hub provides a convenience
method which simulates a synchronous call from the sender's point of view.
The purpose of this is to simplify the use of the protocol
in situations such as scripting environments which cannot easily handle
callbacks.  However, it is RECOMMENDED to use the asynchronous pattern
where possible due to its greater robustness.

\subsection{Extensible Vocabularies}
\label{sect:vocab}

At several places in this document structured information is 
conveyed by use of a controlled but extensible vocabulary.
Some examples are the client metadata keys (Section \ref{sect:app-metadata}),
message encoding keys (Section \ref{sect:msg-encode}) and 
MType names (Section \ref{sect:mtypes}).

Wherever this pattern is used, the following rules apply.
This document defines certain well-known keys with defined meanings.
These may be OPTIONAL or REQUIRED as documented, but if present
MUST be used by clients and hubs in the way defined here.  
All such well-known keys start with the string ``{\tt samp.}''.

Clients and hubs are however free to introduce and use non-well-known
keys as they see fit.  Any string may be used for such a non-standard key, 
with the restriction that it MUST NOT start with the prefix ``{\tt samp.}''.
The prefix ``{\tt x-samp.}'' has a special meaning as described below.

The general rule is that hubs and clients encountering keys which
they do not understand SHOULD ignore them, propagating them to 
downstream consumers if appropriate.
As far as possible, where new keys are introduced they SHOULD be
such that applications ignoring them will continue to behave
in a sensible way.

Hubs and clients are therefore able to communicate information 
additional to that defined in the current version of this document
without disruption to those which do not understand it.
This extensibility may be of use to applications which have mutual 
private requirements outside the scope of this specification,
or to enable experimentation with new features.
If the SAMP community finds such experiments useful,
future versions of this document may bring such functionality within
the SAMP specification itself by defining new keys in the 
``{\tt samp.}'' namespace.  The ways in which these vocabularies 
are used means that such extensions should be possible with 
minimal upheaval to the existing specification and implementations.

Non-well-known keys (those outside of the ``{\tt samp}'' namespace)
fall into two categories:
those which are candidates for future incorporation into the SAMP
standard as well-known, and those which are not.
If developers are experimenting with keys which they hope or believe
may be incorporated into the SAMP standard as well-known at some time
in the future, they may use the special namespace ``{\tt x-samp}''.
If a future version of the standard does incorporate such a key
as well-known, the prefix is simply changed from ``{\tt x-samp.}''
to ``{\tt samp.}''.
Consumers of such keys SHOULD treat keys which differ only in the
substitution of the prefix ``{\tt samp.}'' for ``{\tt x-samp.}'' or
vice versa as if they have identical semantics,
so for instance a client application should treat the value of
a metadata item with key ``{\tt x-samp.a.b}''
 in exactly the same way as one with key ``{\tt samp.a.b}''.
The ``{\tt samp}'' and ``{\tt x-samp}'' form of the same key
SHOULD NOT be presented in the same map.  If both are presented together,
the ``{\tt samp}'' form MAY be considered to take precedence,
though any reasonable behaviour is permitted.
This scheme makes it easy to introduce new well-known keys
in a way which neither makes illicit use of the reserved ``{\tt samp.}''
namespace nor requires frequent updates to the SAMP standard,
and which places a minimum burden on application developers.
Lists of keys in the ``{\tt x-samp}'' namespace under discussion
may be found near \url{http://www.ivoa.net/samp/}.

\subsection{Use of Profiles}
\label{sect:profiles}

The design of SAMP is based on the abstract interfaces
defined in Section~\ref{sect:apis}.  On its own however, this does
not include the detailed instructions required by application developers
to achieve interoperability.  To achieve that, application developers
must know how to map the operations in the abstract SAMP interfaces 
to specific I/O
(in most cases, network) operations.  It is these I/O operations
which actually form the communication between applications.
The rules defining this mapping from interface to I/O operations
are what constitute a SAMP ``Profile''
(the term ``Implementation'' was considered for this purpose, but rejected
because it has too many overlapping meanings in this context).

There are two ways in which such a Profile can be specified as far as
client application developers are concerned:
\begin{enumerate}
\item By describing exactly what bytes are to be sent using what wire
      protocols for each SAMP interface operation
\item By providing one or more language-specific libraries with calls 
      which correspond to those of the SAMP interface
\end{enumerate}
Although either is possible, SAMP is well-suited for approach (1) above
given a suitable low-level transport library.
This is the case since the operations are quite low-level, 
so client applications can easily
perform them without requiring an independently developed SAMP library.
This has the additional advantages that central effort does not have to
be expended in producing language-specific libraries, and that
the question of ``unsupported'' languages does not arise.

Splitting the abstract interface and Profile descriptions 
in this way
separates the basic design principles from the details
of how to apply them, and it opens the door for other Profiles
serving other use cases in the future. 

This document defines two profiles along the lines of (1) above.
The Standard Profile (Section \ref{sect:profile})
which dates from the first version of this document,
is suitable for desktop applications,
while the Web Profile (Section \ref{sect:webprofile}),
introduced at SAMP version 1.3,
is suitable for web (browser-based) applications.

A client author will usually only need to implement SAMP communications
using a single profile.
Hub implementations should ideally implement all known profiles;
in this way clients using different profiles can communicate
transparently with each other via a hub which mediates between them.
Since the different profiles are based on the same abstract interface
(Section \ref{sect:apis}),
such mediation will not lead to loss or distortion of the communications.

\subsection{Security Considerations}
\label{sect:security}

SAMP enables inter-process communications including the capability for
one client to cause execution of code by another client.
This raises the possibility of an unprivileged client performing
privileged actions in virtue of its SAMP-enabled interoperation.
Whether this is problematic in practice depends on two things:
first the identities of the interoperating clients
(whether they all share similar levels of privilege or trust)
and second the semantics of the messages
(the nature of the code that may be executed remotely, and particularly
how it can be parameterised).
In the case that untrusted clients can cause execution of potentially
damaging code by trusted clients, there is a serious security issue.

The trustedness of registered clients is determined by the profile
or profiles operated by the hub at a given time (Section \ref{sect:profiles}),
since the extent to which registered clients are trusted may differ
between different profiles.
Clients registering via the Standard Profile in its usual configuration
can be assumed all to be owned by the same user and hence to have
the same privileges (Section \ref{sect:std-security}),
but Web Profile clients usually have only
limited access privileges outside of the interoperability granted by SAMP
(Section \ref{sect:web-security}).

In most cases profiles will, in virtue of their definition or at least
of their implementation, provide reasonable assurance that
registered clients are unlikely to be hostile.
However for clients which may be run in a general SAMP context,
it is wise not to expose via SAMP mechanisms unrestricted access
to sensitive resources.
In particular, it is recommended not to introduce MTypes which can
be made to execute arbitrary code (inviting injection attacks),
or to declare metadata which reveals sensitive information.
As an alternative approach, it may be appropriate in certain usage
scenarios to ensure that only a restricted secure profile is running.

\section{Abstract APIs and Data Types}
\label{sect:apis}

\subsection{Hub Discovery Mechanism}
\label{sect:hub-discovery}

In order to keep track of which hub is running, a hub discovery mechanism, capable 
of yielding information about how to determine the existence of and communicate with 
a running hub, is needed. This is a Profile-specific matter and
specific prescriptions are described in Sections
\ref{sect:lockfile} (Standard Profile) and \ref{sect:web-httpd} (Web Profile).

\subsection{Communicating with the Hub}
The details of how a client communicates with the hub are Profile-specific.
Specific prescriptions are described in Sections 
\ref{sect:profile} (Standard Profile) and 
\ref{sect:webprofile} (Web Profile).

\subsection{SAMP Data Types}
\label{sect:samp-data-types}

For all hub/client communication, including the actual content of messages,
SAMP uses three conceptual data types:
\begin{enumerate}
\item {\tt string} --- a scalar value consisting of a sequence of characters;
       each character is an ASCII character with hex code
       09, 0a, 0d or 20--7f
\item {\tt list} --- an ordered array of data items
\item {\tt map} --- an unordered associative array of key-value pairs,
      in which each key is a {\tt string} and each value is a data item
\end{enumerate}
These types can in principle be nested to any level, so that the elements
of a list or the values of a map may themselves be strings, lists or maps.

There is no reserved representation for a null value, and it is 
illegal to send a null value in a SAMP context even if the underlying
transport protocol permits this.  However a zero-length
string or an empty list or map may, where appropriate, be used to 
indicate an empty value.

Although SAMP imposes no maximum on the length of a string,
particular transport protocols or implementation considerations 
may effectively do so; in general, hub and client implementations 
are not expected to deal with data items of unlimited size.
General purpose MTypes SHOULD therefore be specified so that
bulk data is not sent within the message or response.  In general it is
preferred to define a message parameter or result element 
as the URL or filename of a potentially
large file rather than as the inline text of the file itself.
SAMP defines no formal list of which URL protocols are permitted
in such cases, but clients which need to dereference such URLs
SHOULD be capable of dealing with at least the ``http'' and ``file'' schemes.
``https'', ``ftp'' and other schemes are also permitted, but when 
sending such URLs, consideration should be given to whether receiving
clients are likely to be able to dereference them.

At the protocol level there is no provision for typing of scalars.
Unlike many Remote Procedure Call (RPC) protocols SAMP does not distinguish syntactically
between strings, integers, floating point values, booleans etc.
This minimizes the restrictions on what underlying transport
protocols may be used, and avoids a number of problems associated with
using typed values from weakly-typed languages such as 
Python and Perl.
The practical requirement to transmit these types is addressed however
by the next section.

\subsection{Scalar Type Encoding Conventions}
\label{sect:scalar-types}

Although the protocol itself defines {\tt string} as the only scalar type,
some MTypes will wish to define parameters or return
values which have non-string semantics,
so conventions for encoding these as {\tt string}s
are in practice required. 
Such conventions only need to be understood by the sender and 
recipient of a given message and so can be established on a per-MType basis, 
but to avoid unnecessary duplication of effort
this section defines some commonly-used
type encoding conventions.

We define the following BNF productions:
\begin{verbatim}
  <digit>        ::= "0" | "1" | "2" | "3" | "4" | "5" | "6"
                   | "7" | "8" | "9"
  <digits>       ::= <digit> | <digits> <digit>
  <float-digits> ::= <digits> | <digits> "." | "." <digits>
                   | <digits> "." <digits>
  <sign>         ::= "+" | "-"
\end{verbatim}
With reference to the above we define the following type encoding conventions:
\begin{itemize}
\item \verb,<SAMP int> ::= [ <sign> ] <digits>,\\
   An integer value is encoded using its decimal representation with
   an OPTIONAL preceding sign and with no
   leading, trailing or embedded whitespace.
   There is no guarantee about the largest or smallest values which can
   be represented, since this will depend on the processing environment
   at decode time. \\
\item \verb,<SAMP float> ::= [ <sign> ] <float-digits>,\\
      \verb,                 [ "e" | "E" [ <sign> ] <digits> ],\\
   A floating point value is encoded as a mantissa with an OPTIONAL
   preceding sign followed by an OPTIONAL exponent part
   introduced with the character ``{\tt e}'' or ``{\tt E}''.
   There is no guarantee about the largest or smallest values which can
   be represented or about the number of digits of precision which are
   significant, since these will depend on the processing environment
   at decode time. \\
\item \verb,<SAMP boolean> ::= "0" | "1",\\
   A boolean value is represented as an integer: zero represents false,
   and any other value represents true.
   1 is the RECOMMENDED value to represent true.
\end{itemize}

The numeric types are based on the syntax of the C programming language,
since this syntax forms the basis for typed data syntax in many other
languages.
There may be extensions to this list in future versions of this standard.

Particular MType definitions may use these conventions or devise
their own as required.  Where the conventions in this list are used, 
message documentation SHOULD make it clear using a form of
words along the lines ``this parameter contains a {\em SAMP int\/}''.

\subsection{Registering with the Hub}
\label{sect:registration}

A client registers with the hub to:
\begin{enumerate} 
    \item establish communication with the hub
    \item advertise its presence to the hub and to other clients
    \item obtain registration information
\end{enumerate}
The registration information is in the form of a {\tt map} containing
data items which the client may wish to use during the SAMP session.
The hub MUST fill in values for the following keys in the returned {\tt map}:
\begin{description}
\item[{\tt samp.hub-id}] ---
   The client ID which is used by the hub when it sends messages itself 
   (rather than forwarding them from other senders).
   For instance, this ID will be used when the hub sends the 
   {\tt samp.hub.event.shutdown} message.
\item[{\tt samp.self-id}] ---
   The client ID which identifies the registering client.
\end{description}
These keys form part of an extensible vocabulary as explained in
Section \ref{sect:vocab}.
In most cases a client will not require either of the above IDs for
normal SAMP operation, but they are there for clients which do wish
to know them.
Particular Profiles may require additional entries in this map.

Immediately following registration, the client will typically
perform some or all of the following OPTIONAL operations:
\begin{itemize}
    \item supply the hub with metadata about itself, using the
    {\tt declareMetadata()} call
    \item tell the hub how it wishes the hub to communicate with it,
    if at all (the mechanism for this is profile-dependent, and it may
    be implicit in registration)
    \item inform the hub which MTypes it wishes to subscribe to, using the
    {\tt declareSubscriptions()} call
\end{itemize}

\subsection{Application Metadata}
\label{sect:app-metadata}

A client may store metadata in the form of a {\tt map} of key-value pairs in the hub 
for retrieval by other clients. Typical metadata might be the human-readable 
name of the application, a description and a URL for its icon, but other values 
are permitted. The following keys are defined for well-known metadata items:
\begin{description}
\item[{\tt samp.name}] --- A one word title for the application.
\item[{\tt samp.description.text}] --- A short description of the 
application, in plain text.
\item[{\tt samp.description.html}] --- A description of the application, 
in HTML.
\item[{\tt samp.icon.url}] --- The URL of an icon in png, gif or jpeg format. 
\item[{\tt samp.documentation.url}] --- The URL of a documentation web page.
\end{description}
All of the above are OPTIONAL, but {\tt samp.name} is strongly RECOMMENDED.
These keys form the basis of an extensible vocabulary as explained in
Section \ref{sect:vocab}.

\subsection{MType Subscriptions}
\label{sect:subscription}

As outlined above, an MType is a string which defines the semantics of 
a message.  MTypes have a hierarchical form.  Their syntax is given
by the following BNF:
\begin{verbatim}
  <mchar>  ::= [0-9A-Za-z] | "-" | "_"
  <atom>   ::= <mchar> | <atom> <mchar>
  <mtype>  ::= <atom> | <mtype> "." <atom>
\end{verbatim}
Examples might be ``{\tt samp.hub.event.shutdown}'' or ``{\tt file.load}''.

A client may {\em subscribe\/} to one or more MTypes to indicate 
which messages it is willing to receive.  A client will only ever
receive messages with MTypes to which it has subscribed.
In order to do this it passes a subscriptions {\tt map} to the hub.
Each key of this map is an MType string to which the client wishes
to subscribe, and the corresponding value is a map which may contain
additional information about that subscription.  Currently, no keys
are defined for these per-MType maps, so typically they will be empty
(have no entries).  The use of a map here is to permit experimentation
and perhaps future extension of the SAMP standard.

As a special case, simple wildcarding is permitted in subscriptions.
The keys of the subscription map may actually be of the form
\verb|<msub>|, where
\begin{verbatim}
  <msub>   ::= "*" | <mtype> "." "*"
\end{verbatim}
Thus a subscription key ``{\tt file.event.*}'' means that a client wishes
to receive any messages with MType which begin ``{\tt file.event.}''.
This does not include ``{\tt file.event}''.
A subscription key ``{\tt *}'' subscribes to all MTypes.
Note that the wildcard ``{\tt *}'' character may only appear
at the end of a subscription key, and that this indicates 
subscription to the entire subtree.

More discussion of MTypes, including their semantics, is given in
Section \ref{sect:mtypes}.

\subsection{Message Encoding}
\label{sect:msg-encode}

A message is an abstract container for the information we wish to send
to another application.  The message itself is that data which should
arrive at the receiving application.  It may be transmitted along
with some external items (e.g.\ sender, recipient and message identifiers)
required to ensure proper delivery or handling.

A message is encoded for SAMP transmission as a {\tt map} with the following REQUIRED keys:
\begin{description}
\item[{\tt samp.mtype}] ---
  A {\tt string} giving the MType which defines the meaning of the message.
  The MType also, via external documentation, defines the names, types and 
  meanings of any parameters and return values.
  MTypes are discussed in more detail in 
  Section \ref{sect:mtypes}.
\item[{\tt samp.params}] ---
  A {\tt map} containing the values for the message's named parameters.
  These give the data required for the receiver to act on the message,
  for instance the URL of a given file.  The names, types and semantics 
  of these parameters are determined by the MType.
  Each key in this map is the name of a parameter, and the corresponding
  value is that parameter's value.
\end{description}
These keys form the basis of an extensible vocabulary as explained in
Section \ref{sect:vocab}.

\subsection{Response Encoding}
\label{sect:response-encode}

A response is what may be returned from a recipient to a sender giving
the result of processing a message (though in the case of the Notification
delivery pattern, no such response is generated or returned).
It may contain MType-specific return values, or error information, or both.

A response is encoded for SAMP transmission as a {\tt map} with the following keys:
\begin{description}
\item[{\tt samp.status}] (REQUIRED) ---
   A {\tt string} summarising the result of the processing.
   It may take one of the following defined values:
   \begin{description}
   \item[{\tt samp.ok}:]
      Processing successful.
      The {\tt samp.result}, but not the {\tt samp.error} entry
      SHOULD be present.
   \item[{\tt samp.warning}:]
      Processing partially successful.
      Both {\tt samp.result} and {\tt samp.error} entries SHOULD be present.
   \item[{\tt samp.error}:]
      Processing failed.
      The {\tt samp.error}, but not the {\tt samp.result} entry
      SHOULD be present.
   \end{description}
   These values form the basis of an extensible vocabulary as explained in
   Section \ref{sect:vocab}.
\item[{\tt samp.result}] (REQUIRED in case of full or partial success) ---
   A {\tt map} containing the values for the message's named return values.
   The names, types and semantics of these returns are determined by
   the MType.
   Each key in this map is the name of a return value, and the corresponding
   value is the actual value.
   Note that even for MTypes which define no return values, the value of this
   entry MUST still be a {\tt map} (typically an empty one).
\item[{\tt samp.error}] (REQUIRED in case of full or partial error) ---
   A {\tt map} containing error information.
   The following keys are defined for this map:
   \begin{description}
      \item[{\tt samp.errortxt}] (REQUIRED) ---
         A short string describing what went wrong.
         This will typically be delivered to the user of the sender application.
      \item[{\tt samp.usertxt}] (OPTIONAL) ---
         A free-form string containing any additional text an application wishes
         to return.  This may be a more verbose error description meant to be  
         appended to the {\tt samp.errortxt} string,
         however it is undefined how this 
         string should be handled when received.
      \item[{\tt samp.debugtxt}] (OPTIONAL) ---
         A longer string which may contain more detail on what went wrong.
         This is typically intended for debugging purposes, and may for instance
         be a stack trace.
      \item[{\tt samp.code}] (OPTIONAL) ---
         A string containing a numeric or textual code identifying the error,
         perhaps intended to be parsable by software.
         Values beginning ``{\tt samp.}'' are reserved.
   \end{description}
   These keys form the basis of an extensible vocabulary as explained in
   Section \ref{sect:vocab}.
\end{description}
These keys form the basis of an extensible vocabulary as explained in
Section \ref{sect:vocab}.

In most cases, such responses will be generated by a Recipient client 
and forwarded by the Hub to the Sender.
In some cases however the hub may pass to the sender an error response it has
generated itself on behalf of the recipient.
In particular, if the hub determines that no response will ever be received
from the recipient (perhaps because the recipient has unregistered
without replying) the hub MAY generate and forward a response with
{\tt samp.status=samp.error} and the {\tt samp.code} key in the 
{\tt samp.error} structure set to ``{\tt samp.noresponse}''.
Clients SHOULD NOT generate such {\tt samp.code=samp.noresponse} responses
themselves.

\subsection{Sending and Receiving Messages}
\label{sect:delivery}

As outlined in Section~\ref{sect:delivery-outline}, 
three messaging patterns are supported, differing
according to whether and how the response is returned to the sender. 
For a given MType 
there may be a messaging pattern that is most typically used, but there is 
nothing in the protocol that ties a particular MType to a particular messaging 
pattern; any MType may legally be sent using any delivery pattern.

From the point of view of the sender, there are three ways in which a message
may be sent,
and from the point of view of the recipient there are two ways in which
one may be received.  These are described as follows.
\begin{description}
\item[Notification:]  In the notification pattern, communication is only
in one direction:
\begin{enumerate}
\item The sender sends a message to the hub for delivery to one or more
   recipients.
\item The hub forwards the message to those requested recipients which are subscribed.
\item No reply from the recipients is expected or possible.
\end{enumerate}
Notifications can be sent to a given recipient or broadcast to all 
recipients.  The notification pattern for a single recipient is illustrated in 
Figure~\ref{fig:notification}.
\begin{figure}[!h]
\begin{center}
\includegraphics[scale=0.5]{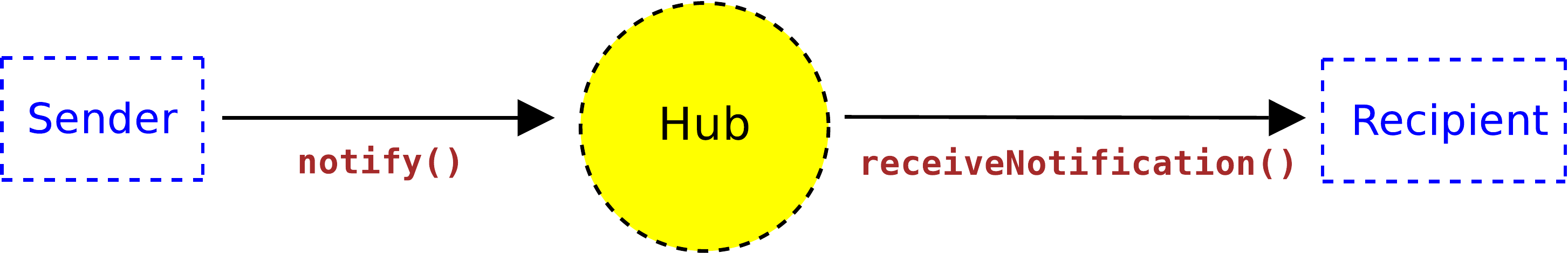}
\caption{Notification pattern}
\label{fig:notification}
\end{center}
\end{figure}
\item[Asynchronous Call/Response:]
In the asynchronous call pattern, {\em message tags\/} and
{\em message identifiers\/} are used 
to tie together messages and their replies:
\begin{enumerate}
\item The sender sends a message to the hub for delivery to one or more
    recipients, supplying along with the message a tag string of its
    own choice, {\em msg-tag\/}.  In return it receives a unique 
    identifier string, {\em msg-id\/}.
\item The hub forwards the message to the appropriate recipients,
    supplying along with the message an identifier string, {\em msg-id\/}.
\item Each recipient processes the message, and sends its response
    back to the hub along with the ID string {\em msg-id\/}.
\item Using a callback, the hub passes the response back to the 
    original sender along with the ID string {\em msg-tag\/}.
\end{enumerate}
The sender is free to use any value for the {\em msg-tag\/}.
There is no requirement on the form of the hub-generated {\em msg-id\/}
(it is not intended to be parsed by the recipient), but it MUST be
sufficient for the hub to pair messages with their responses reliably,
and to pass the correct {\em msg-tag\/} back with the response 
to the sender\footnote{
   One way a hub might implement this is to generate {\em msg-id\/}
   by concatenating the sender's client ID and the {\em msg-tag\/}.
   When any response is received the hub can then unpack the accompanying
   {\em msg-id\/} to find out who the original sender was and what
   {\em msg-tag\/} it used.  In this way the hub can determine
   how to pass each response back to its correct sender without needing
   to maintain internal state concerning messages in progress.
   Hub and client implementations may wish to exploit this freedom 
   in assigning message IDs for other purposes as well, 
   for instance to incorporate timestamps or checksums.
}.
In most cases the sender will not require the {\em msg-id\/}, since
the {\em msg-tag\/} is sufficient to match calls with responses.
For this reason, the sender need not retain the {\em msg-id\/} and
indeed need not wait for it, avoiding a hub round trip at send time.
The only case in which the sender may require the {\em msg-id\/} is
if it needs to communicate later with the recipient about the message
that was sent, for instance as part of a progress report.
Asynchronous calls may be sent to a given recipient or broadcast to all
recipients.  In the latter case, the sender SHOULD be prepared to deal
with multiple responses to the same call.
The asynchronous pattern is illustrated in Figure~\ref{fig:async}.
\begin{figure}[!h]
\begin{center}
\includegraphics[scale=0.45]{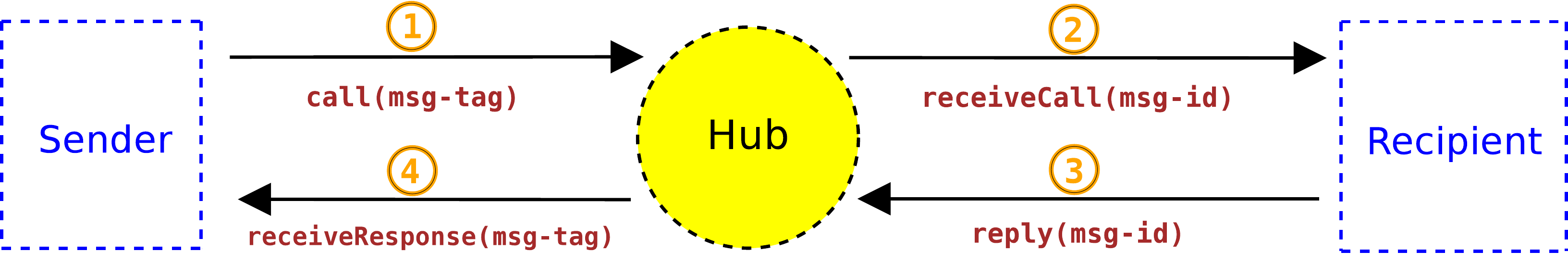}
\caption{Asynchronous Call/Response pattern}
\label{fig:async}
\end{center}
\end{figure}
\item[Synchronous Call/Response]
A synchronous utility method is provided by the hub, mainly for
the convenience of
environments where dealing with asynchronicity might be a problem.
The hub will provide synchronous behaviour to the sender,
interacting with the receiver in exactly the same way as for the
asynchronous case above.
\begin{enumerate}
\item The sender sends a message to the hub for delivery to a given recipient,
      optionally specifying as well a maximum time it is prepared to wait.
      The sender's call blocks until a response is available.
\item The hub forwards the message to the recipient,
    supplying along with the message an ID string, {\em msg-id\/}.
\item The recipient processes the message, and sends its response
    back to the hub along with the ID string {\em msg-id\/}.
\item The hub passes back the response as the return value from the original
    blocking call made by the sender.  If no response is received within
    the sender's specified timeout the blocking call will terminate with 
    an error.  The hub is not guaranteed to wait indefinitely; 
    it MAY in effect impose its own timeout.
\end{enumerate}
There is no broadcast counterpart for the synchronous call.
This pattern is illustrated in Figure~\ref{fig:sync}.
\begin{figure}[!h]
\begin{center}
\includegraphics[scale=0.45]{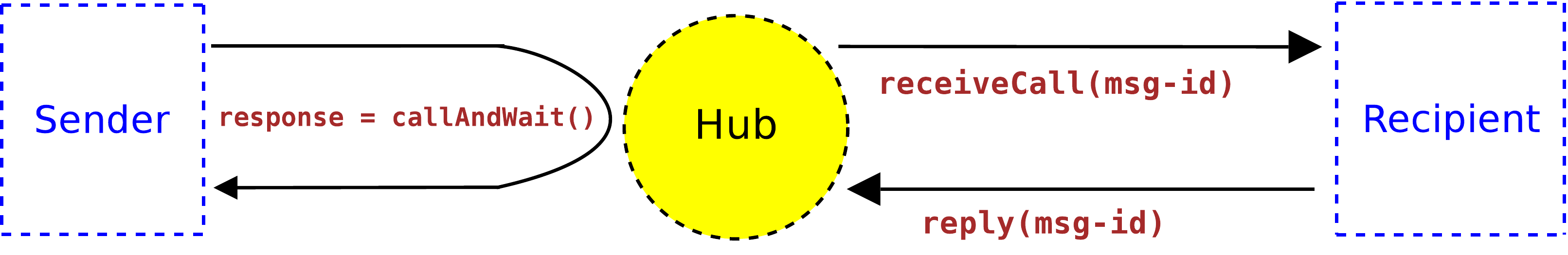}
\caption{Synchronous Call/Response pattern}
\label{fig:sync}
\end{center}
\end{figure}
\end{description}

Note that the two different cases from the receiver's point of view,
{\em Notification\/} and {\em Call/Response\/}, 
differ only in whether a response is returned to the hub.
In other respects the receiver SHOULD process the message in 
exactly the same way for both patterns.

Although it is REQUIRED by this standard that client applications
provide a Response for every Call that they receive, there is no
way that the hub can enforce this.  Senders using the Synchronous
or Asynchronous Call/Response patterns therefore should be aware 
that badly-behaved recipients might fail to respond, leading to calls
going unanswered indefinitely.  The timeout parameter in the
Synchronous Call/Response pattern provides some protection from
this eventuality; users of the Asynchronous Call/Response pattern
may or may not wish to take their own steps.

\subsection{Operations a Hub Must Support}
\label{sect:hubOps}
This section describes the operations that a hub MUST support and the associated
data that MUST be sent and received.
The precise details of how these operations 
map onto method names and signatures is Profile-dependent.
The mapping for the 
Standard Profile is described in Section \ref{sect:mappingXMLRPC},
and for the Web Profile in Section \ref{sect:webXMLRPC}.
\begin{itemize}
    \item \verb|map reg-info = register()|\\
    Method called by a client wishing to register 
    with the hub.
    The form of {\tt reg-info} is given in Section \ref{sect:registration}.
    Note that the form of this call may vary according to the requirements 
    of the particular Profile in use. For instance authentication tokens
    may be passed in one or both directions to complete registration.
    \\
    \item \verb|unregister()|\\
    Method called by a client wishing to unregister from the hub.
    \\
    \item \verb|declareMetadata(map metadata)|\\
    Method called 
    by a client to declare its metadata.
    May be called zero or more times to update hub state; the most recent 
    call is the one which defines the client's currently declared metadata.
    The form of the {\tt metadata} map is given in
    Section \ref{sect:app-metadata}.
    \\
    \item \verb|map metadata = getMetadata(string client-id)|\\
    Returns the metadata
    information for the client whose public ID is {\tt client-id}.
    The form of the {\tt metadata} map is given in 
    Section \ref{sect:app-metadata}.
    \\
    \item \verb|declareSubscriptions(map subscriptions)|\\ 
    Method called by
    a callable client to declare the MTypes it wishes to subscribe to.
    May be called zero or more times to update hub state; the most recent 
    call is the one which defines the client's currently subscribed MTypes.
    The form of the {\tt subscriptions} map is given in
    Section \ref{sect:subscription}.
    \\
    \item \verb|map subscriptions = getSubscriptions(string client-id)|\\
    Returns the subscribed MTypes
    for the client whose public ID is {\tt client-id}.
    The form of the {\tt subscriptions} map is given in
    Section \ref{sect:subscription}.
    \\
    \item \verb|list client-ids = getRegisteredClients()|\\
    Returns the list of public ids of all other registered clients.
    The caller's ID ({\tt samp.self-id} from Section \ref{sect:registration}) 
    is not included,
    but the hub's ID ({\tt samp.hub-id} from Section \ref{sect:registration}) is.
    \\
    \item \verb|map client-subs = getSubscribedClients(string mtype)|\\
    Returns a map with an entry for all other registered clients which are
    subscribed to the MType {\tt mtype}.
    The key for each entry is a subscribed client ID, 
    and the value is a (possibly empty) {\tt map} 
    providing further information on its subscription to {\tt mtype} 
    as described in Section \ref{sect:subscription}.
    An entry for the caller is not included, even if it is subscribed.
    {\tt mtype} MUST NOT include wildcards.
    \\
    \item \verb|notify(string recipient-id, map message)|\\
    Sends a message using the Notification pattern
    to a given recipient.
    The form of the {\tt message} map is given 
    in Section \ref{sect:msg-encode}.
    An error results if the recipient is not subscribed to the message's
    MType.
    \\
    \item \verb|list recipient-ids = notifyAll(map message)|\\ 
    Sends a message using the Notification pattern
    to all other clients which are subscribed to the message's MType.
    The form of the {\tt message} map is given 
    in Section \ref{sect:msg-encode}.
    The return value is a {\tt list} of the client IDs of the clients
    to which an attempt to send the message is made.
    \\
    \item \verb|string msg-id = call(string recipient-id, string msg-tag,|\\
              \verb|                     map message)|\\
    Sends a message using the Asynchronous Call/Response pattern
    to a given recipient.
    The form of the {\tt message} map is given 
    in Section \ref{sect:msg-encode}.
    An error results if the recipient is not subscribed to the message's
    MType, or if the invoking client is not Callable.
    \\
    \item \verb|map calls = callAll(string msg-tag, map message)|\\
    Sends a message using the Asynchronous Call/Response pattern
    to all other clients which are subscribed to the message's MType.
    The form of the {\tt message} map is given 
    in Section \ref{sect:msg-encode}.
    The returned value is a {\tt map} in which the keys are the client IDs
    of clients to which an attempt to send the message is made,
    and the values are the associated {\tt msg-id} strings.
    An error results if the invoking client is not Callable.
    \\
    \item \verb|map response = callAndWait(string recipient-id,|\\
              \verb|                           map message, string timeout)|\\
    Sends a message using the Synchronous Call/Response pattern
    to a given recipient.
    The forms of the {\tt message} and {\tt response} maps are given 
    in Sections \ref{sect:msg-encode} and \ref{sect:response-encode}.
    The {\tt timeout} parameter is interpreted as a {\em SAMP int\/}
    (Section \ref{sect:scalar-types}) giving
    the maximum number of seconds the client wishes to wait.
    If the response takes longer than that to arrive this method SHOULD
    terminate anyway with an error (it MUST not return a {\tt response} indicating
    error).
    Any response arriving from the recipient after that will be discarded.
    If {\tt timeout}$<=0$ then no artificial timeout is imposed.
    An error results if the recipient is not subscribed to the message's
    MType.
    \\
    \item \verb|reply(string msg-id, map response)|\\
    Method
    called by a client to send its response to a given message.
    The form of the {\tt response} map is given 
    in Section \ref{sect:response-encode}.
\end{itemize}

Of these operations, only {\tt callAndWait()} involves blocking communication
with another client.
The others SHOULD be implemented in such a way that clients can expect
them to complete, and where appropriate return a value, on a timescale
short compared to user response time.

\subsection{Operations a Callable Client Must Support}
\label{sect:clientOps}

This section lists the operations which a client MUST support in order
to be classified as callable.
The hub uses these operations when it wishes to 
pass information to a callable client.
Note that callability is OPTIONAL for clients; special (Profile-dependent) steps
may be required for a client to inform the hub how it can be contacted,
and thus become callable.  Clients which are not callable can send
messages using the Notify or Synchronous Call/Response patterns,
but are unable to receive messages or to use Asynchronous Call/Response,
since these operations rely on client callbacks from the hub.

The precise details of how these operations 
map onto method names and signatures is Profile-dependent.
The mapping for 
the Standard Profile is given in Section \ref{sect:mappingXMLRPC}
and for the Web Profile in Section \ref{sect:web-callable}.
\begin{itemize}
    \item \verb|receiveNotification(string sender-id, map message)|\\
    Method called by the hub when dispatching a notification to its
    recipient.
    The form of the {\tt message} map is given 
    in Section \ref{sect:msg-encode}.
    \\
    \item \verb|receiveCall(string sender-id, string msg-id, map message)|\\
    Method called by the hub when dispatching a call to its recipient. The client
    MUST at some later time make a matching call to {\tt reply()} on the hub.
    The form of the {\tt message} map is given 
    in Section \ref{sect:msg-encode}.
    \\
    \item \verb|receiveResponse(string responder-id, string msg-tag,|\\
          \verb|                map response)|\\
    Method used by the hub to dispatch to the sender the response of an
    earlier asynchronous call.
    The form of the {\tt response} map is given 
    in Section \ref{sect:response-encode}.
\end{itemize}

\subsection{Error Processing}
\label{sect:faults}

Errors encountered by clients when processing Call/Response-pattern messages 
themselves (in response to a syntactically legal {\tt receiveCall()}
operation) SHOULD be signalled by returning appropriate content in
the response map sent back in the matching {\tt reply()} call,
as described in Section \ref{sect:response-encode}.

In the case of failed calls of
the operations defined in Sections \ref{sect:hubOps} and \ref{sect:clientOps},
for instance syntactically invalid parameters or communications failures,
hubs and clients SHOULD where possible use the usual error reporting 
mechanisms of the transport protocol in use.

Where it is problematic or impossible to use the transport protocol's
error reporting mechanisms, in the case of a Call/Response pattern message,
the hub MAY signal errors by generating and passing back 
to the sender a suitable response map as described in Section 
\ref{sect:response-encode}.

\section{Standard Profile}
\label{sect:profile}

Section \ref{sect:apis} provides an abstract definition of 
the operations and data structures used for SAMP messaging.
As explained in Section~\ref{sect:profiles}, 
in order to implement this architecture some concrete choices about 
how to instantiate these concepts are required.

This section gives the details of a SAMP Profile based on the 
XML-RPC specification \cite{xmlrpc}.  
Hub discovery is via a lockfile in the user's home directory.

XML-RPC is a simple general purpose Remote Procedure Call
protocol based on sending XML documents using HTTP POST
(it resembles a very lightweight version of SOAP).
Since the mappings from SAMP concepts
such as API calls and data types to their XML-RPC equivalents is very 
straightforward, it is easy for application authors to write 
compliant code without use of any SAMP-specific library code.
An XML-RPC library, while not essential, will make coding much easier;
such libraries are available for many languages.

\subsection{Data Type Mappings}
\label{sect:profile-typemap}

The SAMP argument and return value data types described in
Section \ref{sect:samp-data-types} map straightforwardly onto XML-RPC
data types as follows:
\begin{center}
\begin{tabular}{lcl}
SAMP type    &                       & XML-RPC element \\
\hline
{\tt string} & --- & {\tt <string>} \\
{\tt list}   & --- & {\tt <array>} \\
{\tt map}    & --- & {\tt <struct>} \\
\end{tabular}
\end{center}
The {\tt <value>} children of {\tt <array>} and {\tt <struct>} elements
themselves contain children of type {\tt <string>}, {\tt <array>} or
{\tt <struct>}.

Note that other XML-RPC scalar types ({\tt <i4>}, {\tt <double>} etc)
are not used; even where the semantic sense of a value matches
one of those types it MUST be encoded as an XML-RPC {\tt <string>}.

\subsection{API Mappings}
\label{sect:mappingXMLRPC}

The operation names in the SAMP hub and client abstract APIs
(Sections \ref{sect:hubOps} and \ref{sect:clientOps}) very nearly 
have a one to one mapping with those in the Standard Profile XML-RPC APIs.
The Standard Profile API MUST be implemented as described in Sections
\ref{sect:hubOps} and \ref{sect:clientOps} with the following REQUIRED adjustments:
\begin{enumerate}
\item The XML-RPC method names (i.e.\ the contents of the XML-RPC
      {\tt <methodName>} elements) are formed by 
      prefixing the hub and client abstract API operation names with
      ``{\tt samp.hub.}'' or ``{\tt samp.client.}'' respectively.
\item The {\tt register()} operation takes the following form:
      \begin{itemize}
      \item \verb|map reg-info = register(string samp-secret)|
      \end{itemize}
      The argument is the {\tt samp-secret} value read from the lockfile
      (see Section \ref{sect:lockfile}).
      The returned {\tt reg-info} map contains an additional entry 
      with key {\tt samp.private-key} whose value is a string
      generated by the hub.
\item {\em All\/} other hub and client methods take the 
      {\tt private-key} as their first argument.
\item A new method, {\tt setXmlrpcCallback()} is added to the hub API.
      \begin{itemize}
      \item \verb|setXmlrpcCallback(string private-key, string url)|
      \end{itemize}
      This informs the hub of the XML-RPC endpoint on which the client 
      is listening for calls from the hub.
      The client is not considered Callable unless and until 
      it has invoked this method.
\item Another new method, {\tt ping()} is added to the hub API.
      This may be called by registered or unregistered applications
      (as a special case the {\tt private-key} argument may be omitted),
      and can be used to determine whether the hub is responding to requests.
      Any non-error return indicates that the hub is running.
\end{enumerate}

The {\tt private-key} string referred to above serves two purposes.
First it identifies the client in hub/client communications.
Some such identifier is required, since XML-RPC calls have no other way of
determining the sender's identity.
Second, it prevents application spoofing, since the private key is
never revealed to other applications, so that one application cannot
pose as another in making calls to the hub.

The usual XML-RPC fault mechanism is used to respond to invalid
calls as described in Section \ref{sect:faults}.
The XML-RPC {\tt <fault>}'s {\tt <faultString>} element SHOULD contain a
user-directed message as appropriate and the {\tt <faultCode>} value
has no particular significance.

\subsection{Lockfile and Hub Discovery}
\label{sect:lockfile}

Hub discovery is performed by examining a lockfile to determine
hub connection parameters, specifically the XML-RPC endpoint at
which the hub can be found, and a ``secret'' token which affords
some measure of security, given suitable restrictions on the
lockfile's readability (see Section \ref{sect:std-security}).
To discover the hub, a client must
therefore:
\begin{enumerate}
  \item Determine where to find the lockfile
        (\ref{sect:lockfileLoc})
  \item Read the lockfile to obtain the hub connection parameters
        (\ref{sect:lockfileText})
\end{enumerate}

\subsubsection{Lockfile Location}
\label{sect:lockfileLoc}

The default location of the lockfile is the file named ``{\tt .samp}''
in the user's home directory.  However the content of the environment
variable named SAMP\_HUB can be used to override this default.

The value of the SAMP\_HUB environment variable is of the form
\verb|<samphub-value>|, as defined by the following BNF production:
\begin{verbatim}
   <samphub-value>  ::=  <hub-location>
   <hub-location>   ::=  <stdlock-prefix> <stdlock-url>
   <lockurl-prefix> ::=  "std-lockurl:"
   <stdlock-url>    ::=  (any URL)
\end{verbatim}
The \verb|<stdlock-url>| will typically, but not necessarily, be a
file-type URL (as described in RFC 1738, section 3.10 \cite{rfc1738}).
So for instance to indicate that the
lockfile to be used will be the file ``{\tt /tmp/samp1}'', you would set
\begin{verbatim}
   SAMP_HUB=std-lockurl:file:///tmp/samp1
\end{verbatim}

Although no other form of the \verb|<hub-location>| value is defined
here, the intention is that the SAMP\_HUB environment variable MAY
be used with prefixes other than ``{\tt std-lockurl:}'' to indicate
use of other, non-Standard, profiles.
Issues may in future arise related to the need to indicate multiple
profiles or profile variants at once; the impact of this requirement
on the syntax and semantics of the SAMP\_HUB variable is for now
deferred.

To locate the lockfile therefore, a Standard Profile-compliant client MUST
determine whether an environment variable named SAMP\_HUB exists;
if so, the client MUST examine the variable's value;
if the value begins with the prefix ``{\tt std-lockurl:}''
the client MUST interpret the remainder of the value as a URL
whose content is the text of the lockfile to be used for hub discovery.
If no SAMP\_HUB environment variable exists, the client MUST
use the file ``.samp'' in the user's home directory as the
lockfile to be used for hub discovery.
If the variable exists, but its value begins with a different prefix,
the client MAY interpret that in some non-Standard way for hub discovery.

Rules for a Standard Profile-compliant hub to use when writing lockfiles 
are similar, but if a hub is unable or unwilling to write a lockfile 
such that it can be read using the above procedure, it MUST signal
an error at the startup and then abort.  For practical reasons, a hub
will probably only be able to write a lockfile indicated by a {\tt file}-type
URL, not for instance an arbitrary {\tt http}-type one.
Lockfiles SHOULD be created with appropriate access restrictions
as discussed in Section \ref{sect:std-security}.

The existence or readability of the lockfile MAY be taken 
(e.g.\ by a hub deciding whether to start or not) 
to indicate that a hub is running.
However it is RECOMMENDED to attempt to contact the hub
at the given XML-RPC URL (e.g.\ by calling {\tt ping()})
to determine whether it is actually alive.

The ``home directory'' referred to above is a somewhat system-dependent 
concept: we define it as the value of the {\tt HOME} environment variable on
Unix-like systems and as the value of the {\tt USERPROFILE} environment
variable on Microsoft Windows\footnote{
   Note to Java developers: contrary to what you might expect,
   the {\tt user.home} system property on Windows does {\em not} give you the
   value of {\tt USERPROFILE}.
   See \url{http://bugs.sun.com/bugdatabase/view_bug.do?bug_id=4787931}.
}.  
``Environment variable'' is itself potentially a system-dependent concept,
but it is clear how to interpret it for all platforms on which we 
currently expect SAMP to be used, so no further explanation is provided
here.

In version 1.11 of the standard, the lockfile was always in the
``{\tt .samp}'' file in the user's home directory.
The option of setting the SAMP\_HUB environment variable to override
this has been introduced to allow more flexibility;
for instance one user can run multiple unconnected hubs,
or multiple users can share the same hub.
If no SAMP\_HUB environment variable is defined, client and hub 
behaviour is exactly as in version 1.11.

\subsubsection{Security Considerations}
\label{sect:std-security}

The hub SHOULD normally create the lockfile with file permissions which allow
only its owner to read it.  This provides a measure of security in
that only processes with the same privileges as the hub process
(hence presumably running under the same user ID) will be able to
register with the hub, since only they will be able to provide the
secret token, obtained from the lockfile, which is required for registration.
Thus under normal circumstances all Standard Profile clients can be
presumed to be running with the same level of trust, so that
no security issues of the type discussed in Section \ref{sect:security}
arise.

If the lockfile is made available in some way other than an
owner-only readable file, for instance via
an unprotected {\tt http}-type URL in order to facilitate use of
the same hub by multiple users on different hosts, there is a potential
security risk.  In that case, protection through an authentication and/or
authorization mechanism might be adopted by the hub implementations, for
instance exploiting the TLS cryptographic protocol \cite{rfc2246}.

\subsubsection{Lockfile Content}
\label{sect:lockfileText}

The format of the lockfile is given by the following BNF productions:
\begin{verbatim}
   <file>          ::=   <lines>
   <lines>         ::=   <line> | <lines> <line>
   <line>          ::=   <line-content> <EOL> | <EOL>
   <line-content>  ::=   <comment> | <assignment>
   <comment>       ::=   "#" <any-string>
   <assignment>    ::=   <name> "=" <any-string>
   <name>          ::=   <token-string>
   <token-string>  ::=   <token-char> | <token-string> <token-char>
   <any-string>    ::=   <any-char> | <any-string> <any-char>
   <EOL>           ::=   "\r" | "\n" | "\r" "\n"
   <token-char>    ::=   [a-zA-Z0-9] | "-" | "_" | "."
   <any-char>      ::=   [\x20-\x7f]
\end{verbatim}
The only parts which are significant to SAMP clients/hubs are
(a) existence of the file and (b) {\tt <assignment>} lines.

A legal lockfile MUST provide (in any order) unique assignments for the
following tokens:
\begin{description}
\item[{\tt samp.secret}] ---
      An opaque text string which must be passed to the hub to permit
      registration.
\item[{\tt samp.hub.xmlrpc.url}] ---
      The XML-RPC endpoint for communication with the hub.
\item[{\tt samp.profile.version}] ---
      The version of the SAMP Standard Profile implemented by the hub
      (``\sampversion'' for the version described by this document).
\end{description}
These keys form the basis of an extensible vocabulary as explained in
Section \ref{sect:vocab}.
Other blank, comment or assignment lines may be included as desired.

An example lockfile might therefore look like this:
\begin{quote}\tt
   \# SAMP lockfile written 2011-12-22T05:30:01 \\
   \# Required keys: \\
   samp.secret=734144fdaab8400a1ec2 \\
   samp.hub.xmlrpc.url=http://andromeda.star.bris.ac.uk:8001/xmlrpc \\
   samp.profile.version=\sampversion \\
   \# Info stored by hub for some private reason: \\
   com.yoyodyne.hubid=c80995f1
\end{quote}

\subsubsection{Hub Discovery Sequences}

The hub discovery sequences are therefore as follows:

\begin{itemize}
\item Client startup:
   \begin{itemize}
   \item Determine hub existence as above
   \item If no hub, client MAY start its own hub
   \item Acquire {\tt samp.secret} value from lockfile
   \item If pre-existing or own hub is running, call {\tt register()} and
         zero or more of {\tt setXmlrpcCallback()}, {\tt declareMetadata()},
         {\tt declareSubscriptions()}
   \end{itemize}
\item Hub startup:
   \begin{itemize}
   \item Determine hub existence as above
   \item If hub is running, exit
   \item Otherwise, start up XML-RPC server
   \item Write lockfile containing mandatory assignments including
         XML-RPC endpoint, using appropriate access restrictions
   \end{itemize}
\item Hub shutdown:
   \begin{itemize}
   \item Remove lockfile (it is RECOMMENDED to first check that this
         is the lockfile written by self)
   \item Notify candidate clients that shutdown will occur
   \item Shut down services
   \end{itemize}
\end{itemize}

A hub implementation SHOULD make its best effort to perform the
shutdown sequence above even if it terminates as a result of some
error condition.

Note that manipulation of a file is not atomic, so that race conditions
are possible.  For instance a client or hub examining the lockfile
may read it after it has been created but before it has been populated
with the mandatory assignments, or two hubs may look for a lockfile
simultaneously, not find one, and both decide that they should
therefore start up, one presumably overwriting the other's lockfile.
Hub and client implementations should be aware of such possibilities,
but may not be able to guarantee to avoid them or their consequences.
In general this is the sort of risk that SAMP and its
Standard Profile are prepared to take --- an eventuality which will occur
sufficiently infrequently that it is not worth significant
additional complexity to avoid.
In the worst case a SAMP session may fail in some way, and will have
to be restarted manually.

\subsection{Examples}

Here is an example in pseudo-code of how an application might locate and
register with a hub, and send a message requiring no response to other
registered clients.
{\examplesize
\begin{verbatim}
   # Locate and read the lockfile.
   string hubvar-value = readEnvironmentVariable("SAMP_HUB");
   string lock-location = getLockfileLocation(hubvar-value);
   map lock-info = readLockfile(lock-location);

   # Extract information from lockfile to locate and register with hub.
   string hub-url = lock-info.getValue("samp.hub.xmlprc.url");
   string samp-secret = lock-info.getValue("samp.secret");

   # Establish XML-RPC connection with hub
   # (uses some generic XML-RPC library)
   xmlrpcServer hub = xmlrpcConnect(hub-url);

   # Register with hub.
   map reg-info = hub.xmlrpcCall("samp.hub.register", samp-secret);
   string private-key = reg-info.getValue("samp.private-key");

   # Store metadata in hub for use by other applications.
   map metadata = ("samp.name" -> "dummy",
                   "samp.description.text" -> "Test Application",
                   "dummy.version" -> "0.1-3");
   hub.xmlrpcCall("samp.hub.declareMetadata", private-key, metadata);

   # Send a message requesting file load to all other 
   # registered clients, not wanting any response.
   map loadParams = ("filename" -> "/tmp/foo.bar");
   map loadMsg = ("samp.mtype" -> "file.load",
                  "samp.params" -> loadParams);
   hub.xmlrpcCall("samp.hub.notifyAll", private-key, loadMsg);

   # Unregister
   hub.xmlrpcCall("samp.hub.unregister", private-key);
\end{verbatim}
}

The first few XML-RPC documents sent over the wire for this exchange 
would look something like the following.  
The registration call from the client to the hub:
{\examplesize
\begin{verbatim}
   POST /xmlrpc HTTP/1.0
   User-Agent: Java/1.5.0_10
   Content-Type: text/xml
   Content-Length: 189

   <?xml version="1.0"?>
   <methodCall>
     <methodName>samp.hub.register</methodName>
     <params>
       <param><value><string>734144fdaab8400a1ec2</string></value></param>
     </params>
   </methodCall>
\end{verbatim}
}
which leads to the response:
{\examplesize
\begin{verbatim}
   HTTP/1.1 200 OK
   Connection: close
   Content-Type: text/xml
   Content-Length: 464

   <?xml version="1.0"?>
   <methodResponse>
     <params><param><value><struct>
       <member>
         <name>samp.private-key</name>
         <value><string>client-key:1a52fdf</string></value>
       </member>
       <member>
         <name>samp.hub-id</name>
         <value><string>client-id:0</string></value>
       </member>
       <member>
         <name>samp.self-id</name>
         <value><string>client-id:4</string></value>
       </member>
     </struct></value></param></params>
   </methodResponse>
\end{verbatim}
}
The client can then declare its metadata:
the response to this call has no useful content so can be ignored or discarded.
{\examplesize
\begin{verbatim}
   POST /xmlrpc HTTP/1.0
   User-Agent: Java/1.5.0_10
   Content-Type: text/xml
   Content-Length: 600

   <?xml version="1.0"?>
   <methodCall>
     <methodName>samp.hub.declareMetadata</methodName>
     <params>
       <param><value><string>app-id:1a52fdf-2</string></value></param>
       <param><value><struct>
         <member>
           <name>samp.name</name>
           <value><string>dummy</string></value>
         </member>
         <member>
           <name>samp.description.text</name>
           <value><string>Test application</string></value>
         </member>
         <member>
           <name>dummy.version</name>
           <value><string>0.1-3</string></value>
         </member>
       </struct></value></param>
     </params>
   </methodCall>
\end{verbatim}
}
The message itself is sent from the client to the hub as follows:
{\examplesize
\begin{verbatim}
   POST /xmlrpc HTTP/1.0
   User-Agent: Java/1.5.0_10
   Content-Type: text/xml
   Content-Length: 523

   <?xml version="1.0"?>
   <methodCall>
     <methodName>samp.hub.notifyAll</methodName>
     <params>
       <param><value><string>app-id:1a52fdf-2</string></value></param>
       <param><value><struct>
         <member>
           <name>samp.mtype</name>
           <value>file.load</value>
         </member>
         <member>
           <name>samp.params</name>
           <value><struct>
             <name>filename</name>
             <value>/tmp/foo.bar</value>
           </struct></value>
         </member>
       </struct></value></param>
     </params>
   </methodCall>
\end{verbatim}
}
Again, there is no interesting response.

\section{Web Profile}
\label{sect:webprofile}

This section defines the SAMP Web Profile
which allows web applications
to communicate with a SAMP hub.
A {\em web application\/} in this context is code which is downloaded
by a web browser from a remote server, usually as part of a web page,
and which then runs from within that browser.
The most common platforms (browser-based runtime environments) for such
applications are currently JavaScript (a.k.a.\ JScript, ECMAScript),
Java applets, Adobe Flash, and Microsoft Silverlight.
For security reasons, these runtime environments run the web applications
that they host inside a secure ``sandbox'', which imposes restrictions
on access to resources, making it impossible to use the Standard Profile
defined in Section \ref{sect:profile}.
Java applets provide a client-controlled cross-browser mechanism, 
based on code signing, for circumventing these restrictions,
but the others do not.

Section \ref{sect:web-overview} gives an illustrative overview of the
way the Web Profile achieves its communication requirements,
with comparison to the Standard Profile.
Section \ref{sect:web-hub} describes in detail how the Web Profile hub
is implemented in order to provide the functionality defined by
the SAMP abstract hub and client APIs
(Sections \ref{sect:hubOps} and \ref{sect:clientOps}).
Section \ref{sect:web-client} outlines the steps that a Web Profile client
must take to locate and communicate with the hub.
The important topic of the security implications of this scheme,
and measures which hub implementations can take in view of these,
is covered separately in Section \ref{sect:web-security}.

\subsection{Overview and Comparison with Standard Profile}
\label{sect:web-overview}

The Web Profile is based on the Standard Profile
(Section \ref{sect:profile}), but with some
modifications which allow clients to overcome
the restrictions imposed by the browser sandbox.

Browser restrictions present four main problems for a web-based SAMP client:
hub discovery, outward hub communication, inward hub communication and
use of third-party URLs.
These are solved in the Web Profile by use of a well-known port,
use of standard and de facto cross-origin access techniques,
reversed HTTP communication,
and URL proxying.
These solutions are described, with comparison to the approaches
used by the Standard Profile, in the following subsections.

\subsubsection{Hub Discovery}

A Standard Profile client locates the hub by reading a ``lockfile'' 
at a well-known location in the filesystem, which provides the HTTP
endpoint at which the hub XML-RPC server is listening and a token
which the client must present in order to register.
Web applications have no access to the local filesystem and so
are unable to read such a lockfile.

In the Web profile, the hub HTTP server listens instead
on a well-known port on the local host.
The hub will apply some security measures at registration time
(Section \ref{sect:web-sec-reg}),
but they are not based on presentation of a secret token.

Note that since this well-known port number is fixed,
it is not possible for more than one Web Profile hub to run on
the same host.  The Web Profile Hub and corresponding web browser
MUST run on the same host, and SHOULD always be run by the same user.

For a web client to be able to access this well-known port at all,
the cross-origin techniques discussed in the next section are required.

\subsubsection{Outward Communications}

In the Standard Profile, all hub communication is done using the HTTP-based
XML-RPC protocol \cite{xmlrpc}, usually to a port on the local host.

This is problematic for web-based clients, since so-called 
``cross-origin'' or ``cross-domain'' policies
enforced by browsers restrict HTTP access under normal circumstances
so that web applications may {\em only\/} make HTTP requests to
URLs at their own {\em Origin\/} \cite{origin}, that is to URLs on the server
from which the web application itself was downloaded.
This deliberately excludes access to a server on the local host,
which is where the SAMP hub is likely to reside.

Since cross-origin access is a common requirement for web-based clients,
and it is not always in conflict with the security concerns of servers,
a number of platform-dependent but widely-used mechanisms have been 
implemented in browser technology
which allow a sandboxed client to talk to an HTTP server
which has explicitly opted in for such cross-origin communications.
A Web Profile hub will implement one or more of these cross-origin
workarounds (Section \ref{sect:web-httpd})
and so permit Web Profile clients running in the 
relevant browser runtime environment(s) to make HTTP requests to itself,
thereby allowing client-to-hub XML-RPC calls.

\subsubsection{Inward Communications}

If it wishes to receive as well as send messages, and also to make
asynchronous calls, a SAMP client must declare itself
{\em Callable\/}, by providing the Hub with a profile-dependent means
to invoke the client API defined in Section \ref{sect:clientOps}.

In the Standard Profile a client declares itself Callable by
providing to the Hub an HTTP endpoint to which the Hub may make
XML-RPC requests.  Thus, the client must itself run a publicly
accessible HTTP server in order to be callable.
Running an HTTP server is typically not within the capabilities
of a web application.

In the Web Profile, hub-to-client communication is effected by reversing
the direction of the XML-RPC calls, and hence of the HTTP requests.
Instead of the client running a server which listens
for incoming messages from the Hub, the Hub maintains a queue
of messages destined for the client, and the client polls
the Hub to find out if any are available.  The client may
either make periodic short-timeout requests to the hub, or
make a long-timeout (``long poll'') request which will
return early if and when one or more messages are available.
This effects inward communications using only the same outward
HTTP capability discussed in the previous section.

\subsubsection{Third-Party URLs}

Although it is not fundamental to the SAMP protocol itself, many
SAMP MTypes are defined in such a way that a receiving client
must retrieve data from a URL external to the SAMP client-hub system in order
to act on them.  For instance the {\tt table.load.votable} MType
has an argument named ``{\tt url}'', whose value is the location of
the VOTable document to be loaded.  Such URLs may point
to the local filesystem, to a server run by the sending client,
or to some other web server internal or external to the host on
which the SAMP communications are taking place.
Similar considerations apply to some of the client metadata items
(Section \ref{sect:app-metadata}), for instance {\tt samp.icon.url}.
In any of these cases, it is likely that a browser-based client 
will be blocked by the browser's cross-origin policy from accessing
the content of the resource in question.

The Web Profile therefore mandates that the Hub must provide to
registered clients a mechanism for translating arbitrary URLs
into cross-origin-accessible URLs with the same content as the
specified resource.  Since a hub must already be providing a
cross-origin capable HTTP service accessible from the web client, 
it can use the same mechanism to operate a service which proxies
external resources in a cross-origin capable way.

\subsection{Hub Behaviour}
\label{sect:web-hub}

This section specifies in detail the services that a SAMP hub must
provide in order to implement the SAMP Web Profile.

The Web Profile is based on client-to-hub XML-RPC calls,
with the hub residing at a well-known port,
and some special measures for allowing cross-origin requests.
In most ways it resembles the Standard Profile (Section \ref{sect:profile}),
but there are some differences.

\subsubsection{Data Type Mappings}

SAMP argument and return value data types are encoded into XML-RPC
exactly as for the Standard Profile (Section \ref{sect:profile-typemap}).

\subsubsection{API Mappings}
\label{sect:webXMLRPC}

The operation names in the SAMP hub API very nearly have a one to one
mapping with those in the Web Profile XML-RPC API.  The Web Profile
Hub API MUST be implemented as described in Section \ref{sect:hubOps},
with a number of REQUIRED adjustments.
These are summarised as follows, and described in more detail later.

\begin{enumerate}
\item The XML-RPC method names (i.e.\ the contents of the XML-RPC
      {\tt <methodName>} elements) are formed by prefixing
      the hub abstract API operation names with ``{\tt samp.webhub.}''.
      For brevity, this prefix is not written in the rest of this document,
      but it is to be understood on all hub API XML-RPC calls.
\item The {\tt register} operation takes the following form
      (Section \ref{sect:web-registration}):
      \begin{itemize}
      \item {\tt map reg-info = register(map identity-info)}
      \end{itemize}
      The {\tt identity-info} is a map containing at least a declared
      application name supplied by the registering application to
      indicate its identity.
\item The {\tt reg-info} map returned from the {\tt register} method
      MUST contain two entries additional to those mandated by the
      hub API (Section \ref{sect:web-registration}):
      \begin{description}
      \item[{\tt samp.private-key}:]
           used as the first argument of all hub API XML-RPC calls
      \item[{\tt samp.url-translator}:]
           used for translation of foreign URLs for cross-origin accessibility
      \end{description}
\item {\em All\/} hub methods other than {\tt register}
      take the {\tt private-key} as their first argument,
      except where otherwise noted ({\tt ping}).
      For brevity, this argument is not written in the rest of this document,
      but it is to be understood on all hub API calls.
\item Two new methods are added to the hub API to support reversed callbacks
      (Section \ref{sect:web-callable}):
      \begin{itemize}
      \item {\tt allowReverseCallbacks(string allow)}
      \item {\tt map pullCallbacks(string timeout)}
      \end{itemize}
\item Another new method is added to the hub API:
      \begin{itemize}
      \item {\tt ping()}
      \end{itemize}
      This may be called by registered or unregistered applications
      (as a special case the {\tt private-key} argument may be omitted),
      and can be used to determine whether the hub is responding to requests.
      Any non-error return indicates that the hub is running.
\end{enumerate}

\subsubsection{Hub HTTP Server}
\label{sect:web-httpd}

Communications are XML-RPC calls \cite{xmlrpc} from the client to the Hub.
XML-RPC works using POSTs to an HTTP server.  The Web Profile hub HTTP
server resides on the well-known port 21012,
so that clients know where to find it on the local host.
The XML-RPC endpoint for Web Profile requests
is at the root of that server, so that web clients can access it by
POSTing to the URL ``{\tt http://localhost:21012/}''.

In general, web applications operate inside a browser-enforced sandbox that
prevents them from accessing cross-origin resources, including HTTP-based ones
served from the local host.  However there are a number of ways in which
an HTTP server can elect to permit access from browser-based clients.
In order to be useful a Web Profile hub must implement at
least one of these ``cross-origin workarounds''.

The following cross-origin workarounds are known to exist, and
can be considered for use by Web Profile hub HTTP servers:
\begin{description}
\item[Cross-Origin Resource Sharing:]
CORS \cite{cors} is a W3C standard which
works by manipulation of the HTTP Origin header and related headers
by the browser runtime environment and the HTTP server,
allowing the HTTP server to grant
cross-domain access from clients with some or all Origins.
CORS forms part of the XmlHttpRequest Level 2 standard \cite{xhr2},
which is implemented by, at least, 
Chrome v2.0+, Firefox v3.5+ and Safari v4.0+.
Microsoft's IE8+ implements CORS via its own non-standard XDomainRequest
object.  
This standard belongs to the loose HTML5 family of technologies,
and it is likely that support will become wider in the future.
A Web Profile hub HTTP server can grant unrestricted access
to CORS-aware web applications
by following the instructions in the CORS standard to enable both 
{\em simple\/} and {\em preflight\/} requests from clients
with any Origin.
\item[Flash cross-domain policy:] Adobe's Flash browser plugin makes
use of a resource named ``{\tt crossdomain.xml}'', which, if present on
an external HTTP server, is taken to indicate willingness to serve
cross-domain requests \cite{flash-crossdomain}.
This has emerged as something of a de facto standard, and the
crossdomain file is honoured by Silverlight and unsigned Java
Applets/WebStart applications\footnote{Support for the crossdomain.xml
   file is reportedly implemented in Java v1.6.0\_10 and later,
   see \url{http://bugs.sun.com/bugdatabase/view_bug.do?bug_id=6676256}.}
as well as for Flash applications.
A Web Profile hub HTTP server can grant unrestricted access
to Flash-like web applications
by serving a resource named ``{\tt /crossdomain.xml}''
with a Content-Type header of
``{\tt text/x-cross-domain-policy}'' and content like:
{\footnotesize\begin{verbatim}
   <?xml version="1.0"?>
   <!DOCTYPE cross-domain-policy
             SYSTEM "http://www.adobe.com/xml/dtds/cross-domain-policy.dtd">
   <cross-domain-policy>
     <site-control permitted-cross-domain-policies="all"/>
     <allow-access-from domain="*"/>
     <allow-http-request-headers-from domain="*" headers="*"/>
   </cross-domain-policy>
\end{verbatim}}
\item[Silverlight cross-domain policy:]
Microsoft's Silverlight environment will take note of Flash-style
{\tt crossdomain.xml} files, so the above measure ought to permit
Silverlight clients to access a compliant HTTP server.  However,
Silverlight has its own cross-domain policy mechanism
\cite{silverlight-crossdomain}, which may be implemented in addition.
A Web Profile hub HTTP server can grant unrestricted access
to Silverlight web applications
by serving a resource named ``{\tt /clientaccesspolicy.xml}''
with a Content-Type header of ``{\tt text/xml}'' and content like:
{\footnotesize\begin{verbatim}
   <?xml version="1.0"?>
   <access-policy>
     <cross-domain-access>
       <policy>
         <allow-from>
           <domain uri="http://*"/>
         </allow-from>
         <grant-to>
           <resource path="/" include-subpaths="true"/>
         </grant-to>
       </policy>
     </cross-domain-access>
   </access-policy>
\end{verbatim}}
\end{description}

If the hub implements these cross-origin workarounds 
it is believed that cross-origin access, hence Web Profile SAMP access,
can be provided from nearly all browsers.
Most modern browsers support CORS for JavaScript, 
nearly all others support Flash,
and it is possible for JavaScript applications to 
make use of Flash libraries for their
SAMP communications\footnote{See for instance the
  flXHR library at \url{http://flxhr.flensed.com/}.
}.
Maximum interoperability therefore can be achieved by 
implementing all of these, or at least CORS and Flash,
in the Web Profile HTTP server.  
There are however security implications of which ones to implement,
discussed in Section \ref{sect:web-sec-confirm}.

In the usual browser-hub configuration,
web applications will always seek the Web Profile HTTP server on the
local host.
Since no legitimate use of the Web Profile HTTP server is expected
from non-local hosts, it is therefore strongly RECOMMENDED for 
security reasons that the Web Profile HTTP server refuses
HTTP requests from external hosts with a 403 Forbidden status.
This recommendation and possible exceptions to it are discussed
further in Section \ref{sect:web-sec-host}.

\subsubsection{Registration}
\label{sect:web-registration}

In order to request registration with the Web Profile, a client needs
to invoke the following XML-RPC method:
\begin{verbatim}
   map register(map identity-info)
\end{verbatim}
The {\tt identity-info} map provides information identifying the
registering application which can inform the
hub's decision about whether to allow registration.
It has the following REQUIRED entry:
\begin{description}
\item[{\tt samp.name}] ---
  A {\tt string} giving the name of the application wishing to register,
  in a form that can be presented to the user.
  This SHOULD be the same as the value of the {\tt samp.name} key in the
  application metadata as described in Section \ref{sect:app-metadata}.
\end{description}
Particular implementations or future versions of this standard
may specify additional required or optional entries to this map.

The hub will accept or reject the registration based on the contents of
the {\tt identity-info} map, available information from the HTTP connection
carrying the XML-RPC call, user confirmation,
and the hub's own security policy, as discussed in \ref{sect:web-sec-reg}.
The {\tt register} XML-RPC request will not return until the
hub has decided whether to accept registration.
This decision may involve user interaction and hence take a significant
amount of time.
The likely timescales mean that an HTTP timeout is possible 
but not very probable; in case of a timeout, registration fails.

If registration is accepted, the hub MUST return to the client
a SAMP map containing the entries mandated by Section \ref{sect:registration}
and also the following entries:

\begin{description}
\item[{\tt samp.private-key}:]
     The value of this key is a string which identifies the registered client.
     This string SHOULD be difficult for third parties to guess.
     This arrangement is the same as for the Standard Profile
     (Section \ref{sect:mappingXMLRPC})
\item[{\tt samp.url-translator}:]
     The value of this key is a string which forms the base for a URL
     proxying service, used as described in Section \ref{sect:web-urltrans}
\end{description}

If registration is rejected, the hub MUST return to the client
an XML-RPC Fault, which SHOULD have a suitably explanatory
{\tt faultString}.

\subsubsection{Callable Clients}
\label{sect:web-callable}

In order to be able to receive communications (incoming messages and
asynchronous call replies) {\em from\/} the hub, the Web Profile
provides for the client to be able to poll the hub server for any
messages or replies which are ready for receipt.
In this way, such communications
are pulled by the client rather than being pushed by the hub, so that
no server component is required on the client side.

Two hub methods are provided to implement this:
\begin{itemize}
\item {\tt allowReverseCallbacks(string allow)}
\item {\tt list pullCallbacks(string timeout-secs)}
\end{itemize}
Both these methods, like the others in the interface, are named with the
{\tt samp.webhub.} prefix and take the {\tt private-key} as an additional
first argument.

The {\tt allow} argument of {\tt allowReverseCallbacks} is a
{\em SAMP boolean\/} (``0'' for false or ``1'' for true), and
the {\tt timeout-secs} argument of {\tt pullCallbacks} is a {\em SAMP int}
(see Section \ref{sect:scalar-types}).

If a client intends at some time in the future to poll for callbacks
it MUST invoke {\tt allowReverseCallbacks} with a true argument.
If at some later point it decides that it will remain registered but
will never poll for callbacks again it SHOULD invoke 
{\tt allowReverseCallbacks} with a false argument
(most clients will never make this second call).
The client becomes {\em Callable\/} only when it has invoked this
method with a true argument.

Having invoked {\tt allowReverseCallbacks} with a true argument,
the client SHOULD periodically invoke {\tt pullCallbacks} whose
return value gives the details of
any callbacks ready for dispatch to the client.
The {\tt timeout-secs} parameter is the maximum number of seconds the
client wishes to wait for a response.  When the method is called,
the hub SHOULD wait until at least one callback is available, and
at that point SHOULD return any pending callbacks.
If the elapsed time since {\tt pullCallbacks} was received
exceeds the number of seconds given by the {\tt timeout-secs} argument,
the hub SHOULD return with an empty list of callbacks.
A client may therefore make a non-waiting poll by using a 
{\tt timeout-secs} argument of 0.
The hub MAY return with an empty list of callbacks before the
given timeout has elapsed, for instance if it reaches an internal
timeout limit.

The hub MAY discard pending messages before they have been polled for by
the client, for instance to avoid excessive usage of resources to
store them.  If a {\tt receiveCall} for an Asynchronous Call/Response-pattern
message is discarded in this way, the hub SHOULD inform the sender by 
passing back a {\tt samp.code=samp.noresponse}-type error response,
as described in Section \ref{sect:response-encode}.

The format of the returned value from {\tt pullCallbacks} is a {\tt list}
of elements each of which is a {\tt map} representing a callback
corresponding to one of the methods in the SAMP client API
(Section \ref{sect:clientOps}).
Each of these callbacks is encoded as a {\tt map} with the
following REQUIRED keys:
\begin{description}
\item[{\tt samp.methodName}] ---
     The client API method name for the callback.
     Its value is a {\tt string} taking one of the values
     ``{\tt receiveNotification}'', ``{\tt receiveCall}'' or
     ``{\tt receiveResponse}''.
\item[{\tt samp.params}] ---
     A {\tt list} of the parameters taken by the client API method in
     question, as documented in Section \ref{sect:clientOps}.
\end{description}
These items correspond to the elements present in an XML-RPC call.

Here is an example of a call to {\tt pullCallbacks}.
The client POSTs an XML-RPC call which requests any callbacks 
which are currently pending or which 
become available during the next 600 seconds:
{\examplesize\begin{verbatim}
POST /
Host: localhost:21012
User-Agent: Mozilla/5.0 (X11; U; Linux i686; en-US; rv:1.9.2.11)
 Gecko/20101028 Red Hat/3.6-2.el5 Firefox/3.6.11
Referer: http://www.star.bris.ac.uk/~mbt/websamp/sample.html
Content-Length: 284
Content-Type: text/plain; charset=UTF-8
Origin: http://www.star.bris.ac.uk

<?xml version='1.0'?>
<methodCall>
  <methodName>samp.webhub.pullCallbacks</methodName>
  <params>
    <param>
      <value><string>wk:1_fjlyrdtwtigfqhnwkqokqpbq</string></value>
    </param>
    <param>
      <value><string>600</string></value>
    </param>
  </params>
</methodCall>
\end{verbatim}}
The response, which is returned by the hub after some delay 
between 0 and 600 seconds, specifies a {\tt receiveCall} operation
that the client should respond to:
{\examplesize\begin{verbatim}
200 OK
Content-Length: 1444
Content-Type: text/xml
Access-Control-Allow-Origin: http://www.star.bris.ac.uk

<?xml version='1.0' encoding='UTF-8'?>
<methodResponse>
  <params>
    <param>
      <value>
        <array>
          <data>
            <value>
              <struct>
                <member>
                  <name>samp.methodName</name>
                  <value>samp.webclient.receiveCall</value>
                </member>
                <member>
                  <name>samp.params</name>
                  <value>
                    <array>
                      <data>
                        <value>hub</value>
                        <value>hub_A_cc55_Ping-tag</value>
                        <value>
                          <struct>
                            <member>
                              <name>samp.mtype</name>
                              <value>samp.app.ping</value>
                            </member>
                            <member>
                              <name>samp.params</name>
                              <value>
                                <struct>
                                </struct>
                              </value>
                            </member>
                          </struct>
                        </value>
                      </data>
                    </array>
                  </value>
                </member>
              </struct>
            </value>
          </data>
        </array>
      </value>
    </param>
  </params>
</methodResponse>
\end{verbatim}}
Some of the HTTP headers in the outgoing request in this example
have been added outside of the client's control by the browser 
runtime environment.
In particular the {\tt Origin} inserted by the browser, and the 
{\tt Access-Control-Allow-Origin} provided in response by the Hub,
indicate that CORS negotiation \cite{cors} is in operation
here to allow cross-origin access.

\subsubsection{URL Translation}
\label{sect:web-urltrans}

In order that sandboxed clients are able to obtain the content of
URLs from foreign domains, as is often required by SAMP interoperation,
the hub provides a service which is able to dereference general URLs.

At registration time, as described in Section \ref{sect:web-registration},
one of the values provided to the registering client is that of the
{\tt samp.url-translator} key.  This is a partial URL which, when
another URL {\em u1\/} is appended to it, will return the same content as
{\em u1\/} from an HTTP GET request.  
If {\em u1} is a syntactically legal URL according to
RFC 2396 \cite{rfc2396}, no additional encoding
needs to be performed on it by the client prior to the concatenation.

A sample of ECMAScript code using this facility might look something
like this:
{\footnotesize\begin{verbatim}
   var url_trans = reg_info["samp.url-translator"];
   var u1 = msg["samp.params"]["url"];   // base URL received from message
   var u2 = url_trans + u1;              // URL ready for retrieval
\end{verbatim}}

The partial translator URL might typically be implemented as a URL
pointing to the same HTTP server in which the hub is hosted, with an
empty query part.  The content of URLs accessed in this way SHOULD be 
available under the same cross-origin arrangements described
in Section \ref{sect:web-httpd}.  For security reasons the hub SHOULD ensure
that this facility can only be used by registered clients, for instance
by embedding the private key in the URL.  Thus a translator URL might
look something like
\begin{quote}
{\tt http://localhost:21012/translator}/{\sl client-private-key\/}{\tt ?}
\end{quote}

The URL translation service SHOULD in general write an HTTP response
with HTTP headers appropriate for the resource being served,
in accordance with the HTTP version in use (e.g.\ \cite{rfc2616}).
Where the content type of
a resource is not known (which is typical if that resource is backed
by a file rather than an HTTP URI) the HTTP Content-Type header MAY
be omitted.

For security reasons, such a hub URL translation service MAY refuse
access to certain resources, as discussed in Section \ref{sect:web-sec-urls}.

\subsection{Client Behaviour}
\label{sect:web-client}

The steps that a client must take to register with a Web Profile hub and
participate in two-way SAMP communications are as follows:
\begin{enumerate}
\item Prepare to make XML-RPC communications with the XML-RPC endpoint
      {\tt http://localhost:21012/}.
      Web applications will need to do this using a client which
      supports one of the cross-origin workarounds described in
      Section \ref{sect:web-httpd} and supported by the Web Profile hub.
\item Call the {\tt register}
      XML-RPC method supplying a short application name
      and possibly other information in the {\tt identity-info} argument.
      If this succeeds (returns a non-Fault XML-RPC response),
      the client is registered.
\item If the client wishes to receive as well as send communications
      (to be {\em Callable\/}), first call
      {\tt allowReverseCallbacks} and then
      periodically call {\tt pullCallbacks}.
      Call {\tt declareSubscriptions} as required.
\item Act on retrieved callbacks as required.
      If any MType argument or return value is a URL,
      prefix it with the value of the {\tt samp.url-translator}
      entry from the registration map before dereferencing it.
\item Send SAMP messages etc as required.
\item Unregister when no further SAMP activity is required,
      either because the user requests disconnection or on
      page unload or a similar event.
\end{enumerate}

\subsection{Security Considerations}
\label{sect:web-security}

Web browsers implement cross-origin access restrictions in order to prevent
web applications from activity on a local host which presents
a security risk, for instance reading and writing local files.
This means that, at least in principle, a user can visit a web page
without worrying about security issues, in a way which is not the
case if they download and install an application to run outside
a browser.

The Web Profile described in the preceding subsections however
relies on neutralising these security measures to some extent.
Although it only affects access to a single resource, the HTTP
server on which the Web Profile hub resides, it is potentially serious
since the services provided by the hub can expose sensitive
resources.

Section \ref{sect:web-sec-analysis} below presents an analysis of the risks,
Sections \ref{sect:web-sec-reg} and \ref{sect:web-sec-behave}
outline how they may be mitigated,
and Section \ref{sect:web-sec-summary} summarises the security status of
Web Profile hub deployments in practice.

\subsubsection{Risk Analysis}
\label{sect:web-sec-analysis}

Implementation in the Web Profile of one or more of the sandbox-defeating 
cross-origin workarounds described in Section \ref{sect:web-httpd}
allows an untrusted, hence potentially hostile, web application
to make HTTP requests to the Web Profile SAMP hub HTTP server.
In the first instance, there is only one potentially sensitive
action that this access permits: attempting to register
with the SAMP hub.  If the registration attempt is denied,
the web application can perform no useful or potentially dangerous
operations (except for a denial of service attack, which sandboxed
web applications are capable of in any case).
If the registration is granted, the client can perform two classes
of sensitive actions: first, exchange SAMP messages with other clients,
and second, use the hub's URL translation service to access
cross-domain URLs which would normally be blocked by the browser.

In order to protect against security breaches related to the
Web Profile therefore, two lines of defence may be established:
first, exercise control over which web applications are
permitted to register, and second, restrict the actions that 
registered applications are permitted to take.
These options are explored in the following sections,
\ref{sect:web-sec-reg} and \ref{sect:web-sec-behave} respectively.

\subsubsection{Registration Restrictions}
\label{sect:web-sec-reg}

A running Web Profile implementation may receive requests to register
from any web application running in a local browser, and even some
clients in other categories.
Since not all such applications may be trustworthy, the Web Profile
SHOULD exercise careful control over which ones are permitted to register.
A Web Profile implementation is permitted to make such decisions
in accordance with whatever security policy it deems appropriate,
but it is RECOMMENDED that at least the restrictions described in
the following subsections are considered:
restricting requests to the local host (Section \ref{sect:web-sec-host}),
requiring explicit user confirmation (Section \ref{sect:web-sec-confirm})
and attempting client authentication (Section \ref{sect:web-sec-auth}).

\paragraph{Local Host Restriction}
\label{sect:web-sec-host}

As strongly RECOMMENDED in Section \ref{sect:web-httpd}, registration requests,
and in fact all access to the hub HTTP server, SHOULD under normal
circumstances only be permitted
from the local host.  This blocks registration attempts from web or
non-web applications on the internet at large.

Given this restriction, the only applications which may attempt to
register with a hub run by user U are therefore:
\begin{enumerate}
\item web applications running in a browser run by user U on the local host
\item non-web applications run by user U on the local host
\item web or non-web applications run by users other than U on the local host
\end{enumerate}
Type 1 are the applications that the Web Profile is designed to serve.
Type 2 are not what the Web Profile is designed for, since they could
use the Standard Profile instead, but they already have user privileges
so present no additional security risk.
Type 3 are potentially problematic, if the host in question is a 
multi-user machine, since they may result in a different user who
is already able to run processes on the local host acquiring access to
the hub-owner's resources (e.g.\ private files).
In practice the User Confirmation step (Section \ref{sect:web-sec-confirm})
should serve to distinguish type 3 from legitimate (type 1) requests,
and the behaviour restrictions described in Section \ref{sect:web-sec-behave}
will limit any potential damage.

There may be circumstances under which it is appropriate to relax this
local host restriction, for instance to enable collaboration with
a known external host not capable of Standard Profile communication,
such as a mobile device operated by the hub user.
However, it is RECOMMENDED that Web Profile implementations at 
least restrict access to the local host in their default 
configuration, and if access is permitted to external hosts
it is only by explicit user request, and to a named host or
list of hosts.
Opening the well-known Web Profile hub server port to the internet at
large would invite denial of service and perhaps phishing attacks
in which the user is exposed to unwanted SAMP registration 
requests.

\paragraph{User Confirmation}
\label{sect:web-sec-confirm}

It is strongly RECOMMENDED that the Hub requires explicit confirmation
from the user before any Web Profile application is allowed to register.
This will normally take the form of the Hub popping up a dialogue window
which requires the user to click ``OK'' or similar for registration to proceed.
An implication of this is that the Web Profile hub must have access to
the same visual display on which the browser is running, which almost
certainly means the hub and the browser are run by the same user.

When enquiring about authorization the hub should make clear to the
user the security implications of accepting the registration request,
and should also present to the user any known information about
the application attempting to register.
Unfortunately, little such information is guaranteed to be available.
The name declared by the application as part of its registration request
will be present, but the application is free to declare any name,
perhaps a misleading one.
Certain HTTP headers on the incoming request may also be relevant:
the ``Origin'' header \cite{origin} will be present for requests
originating from CORS, and
the ``Referer'' header \cite[section 14.36]{rfc2616} may be provided,
though its presence and reliability is dependent on the combination of
browser, platform and cross-origin workaround.
Note that use of non-CORS options might on some browser/plugin
platforms permit faking of HTTP headers\footnote{See for example
   \url{http://secunia.com/advisories/22467/},
   which refers to a Flash version from 2006.
   Hopefully browsers and plugins in current use do
   not contain such vulnerabilities,
   but an assurance of this is beyond the scope of this document.},
so that if the Web Profile HTTP server implements one of the non-CORS
options alongside CORS this may reduce the reliability of
header information even from HTTP requests which (apparently)
originate from CORS.  These headers should therefore be used with care.

Since only the name, which may be chosen at will by the registering
application, is guaranteed present, this looks on the face of it
like a poor basis on which to accept or reject registration by
a potentially hostile web application.

However, in practice the timing of the request presentation provides
the most useful information about the identity and credibility of
the request.  A user will only see such a popup dialogue at the time
that a web application attempts to register with SAMP.
This will normally be immediately following a deliberate user browser 
action like opening, or clicking a ``Register'' button on, a web page.
If the user trusts the web page he has just interacted with, 
he can trust the application within it, and should hence authorize 
registration.  If the user does not trust the web page he has 
just interacted with, or if the popup appears at a time when no 
obvious action has been taken to trigger a SAMP registration,
then the user should deny registration.
This pattern of user interaction, requiring authorization based on
the timing of actions in a browser, is both intuitive and familiar
to users; for instance it is used when launching a signed Java applet
or Java WebStart application.

\paragraph{Client Authentication}
\label{sect:web-sec-auth}

As an additional security measure it would be desirable to make a
reliable identification of the author of a web application
by examining an associated digital certificate, with reference
to a list of trusted certificate authorities.
If a certificate reliably associated with the application could be
obtained, this additional information could be presented
to the user or used automatically by the hub to inform the decision
about whether to accept or reject the registration request.

Unfortunately however the content of the actual application is not
available to the Hub at registration time,
so signing the application code will not in itself help.

The Web Profile does not at present therefore make any recommendation
concerning client authentication.  Implementations may however
wish to attempt some level of authentication, perhaps by somehow 
associating a certificate with the web client's URL or Origin 
using the HTTP (or HTTPS) request headers noted in 
Section \ref{sect:web-sec-confirm}, or by use of additional
credentials passed in the {\tt identity-info} map.

\subsubsection{Behaviour Restrictions}
\label{sect:web-sec-behave}

Given the restrictions on client registration recommended by
Section \ref{sect:web-sec-reg}, there is a reasonable expectation
that clients registered with the Web Profile will be trustworthy.
However, the possibility remains that user carelessness or
some phishing-like attack might lead to registration of hostile
clients, and so Web Profile implementations may additionally restrict
the behaviour of registered clients.
In general, a Web Profile hub implementation MAY impose such restrictions
as it sees fit, based on its chosen security policy.
This may lead to the inability of some Web Profile clients
to perform some legitimate SAMP operations;
in such cases the hub SHOULD signal that fact to the client
using an appropriate error mechanism.

Restrictions may be applied as described in the following subsections:
restricting the MTypes that may be sent (Section \ref{sect:web-sec-mtypes}),
and restricting the scope of the URL translation service
(Section \ref{sect:web-sec-urls}).

\paragraph{MType Restrictions}
\label{sect:web-sec-mtypes}

The SAMP standard imposes no restriction on the semantics of MTypes,
so SAMP can in principle be used to send messages which exercise
the privileges available to other SAMP clients in arbitrary ways.
In practice, most SAMP MTypes are fairly harmless; a typical result
is loading an image into an image viewer.  While hostile
abuse of such a capability could be annoying, it does not consitute
a serious security concern.  However one might imagine an MType
that intentionally or unintentionally allowed execution of 
arbitrary scripting operations within the context of a connected
client, and hostile abuse of such a facility could easily result
in theft of or damage to data, or in other serious security breaches.

With this in mind, Web Profile hub implementations MAY
impose some restrictions on the MTypes that registered clients are
permitted to send, via for instance some per-MType whitelisting
or blacklisting mechanism.  Given the open-ended nature of the MType
vocabulary, a whitelisting approach may be most appropriate.

The hub MAY also restrict MTypes that Web Profile registered clients
are permitted to receive, though it is harder to imagine exploits
based on message receipt.

Hubs may implement such message blocking either by hiding
blocked subscriptions from other clients as appropriate,
or by refusing to forward messages corresponding to blocked subscriptions.
In the latter case a communication failure should be signalled
by responding with an XML-RPC fault.

\paragraph{URL Restrictions}
\label{sect:web-sec-urls}

As explained in Section \ref{sect:web-urltrans}, the Web Profile
provides a service for proxying arbitrary URLs, so that
web clients can access data referenced by URL in SAMP messages
or metadata, which sandbox-imposed cross-origin restrictions would 
otherwise block them from reading.

This capability is essential for worthwhile use of many common
SAMP MTypes.
However, it is also open to abuse, for instance a hostile client 
might request to read {\tt file:///etc/passwd} or some HTTP URL 
on the local host or network which is restricted to local access.

Web Profile implementations therefore MAY impose such restrictions
as they see fit on the use of the URL translation service provided
to web clients, in order to prevent such abuse.
Blocking all access to resources which are local
({\tt file:} or {\tt http://localhost/}) is too strict to be useful,
since the URLs referenced in SAMP messages very often
fall into this category.

An appropriate policy might be to proxy only
URLs which a web client is known to have some legitimate
SAMP-based reason to access, namely those which have previously appeared
in the metadata declared by,
or in a message or response originating from, some other client.
In consideration of the fact that web clients might be able to
provoke other clients to emit a chosen URL, or might cooperate
between themselves, such a list of permitted values SHOULD 
be further restricted to those URLs which first appeared
in a metadata or message content or response map from a trusted 
(i.e.\ non-web) client.

Since the hub in general lacks the relevant semantic knowledge
there is no foolproof way to identify URLs in metadata or messages,
but checking for syntactically suitable map values
(e.g. {\tt (http|https|ftp|file)://.*})
is likely to be good enough for this purpose.

Where the Web Profile implementation declines a given URL proxy
request, it MUST respond with a 403 Forbidden HTTP response.

It is also RECOMMENDED that proxied HTTP access is limited to
the ``safe'' HTTP methods GET and optionally HEAD
\cite[section 9.1.1]{rfc2616},
and that user credentials (cookies, authentication etc) are not propagated.
Requests using unsupported HTTP methods MUST be met with a
405 Method Not Allowed response.

\subsubsection{Security Summary}
\label{sect:web-sec-summary}

The basic mechanics of the Web Profile present significant security risks
for a host on which it runs.  This section has described how
security-conscious implementations of the Profile can mitigate those risks.
Following the recommendations from Section \ref{sect:web-sec-reg}
on when to permit registration provides a reasonable assurance that
registered clients will be trustworthy, and in particular guarantees
that clients can only register with explicit authorization from a 
human user.
Following the recommendations from Section \ref{sect:web-sec-behave}
about permitted behaviour of registered clients
ensures that even if a hostile client is allowed to register
it is unlikely to be able to do significant damage.
By combining these measures it is believed that the level of risk
associated with running a Web Profile, while it would not be appropriate
for instance for financial transactions, is no greater than that
encountered on a regular basis by use of the web in general.

The mitigation measures are presented as (in some cases strong)
RECOMMENDations and suggestions rather than REQUIREments,
in order to allow implementations to
experiment with the most appropriate configurations, which may change
as a result of emerging technology and common usage patterns.
Such experimentation and further consideration may result in some 
modification of the protocol or documentation of best practice in
future versions of this document or elsewhere.

\section{MTypes: Message Semantics and Vocabulary}
\label{sect:mtypes}

        A message contains an MType string that
defines the semantic meaning of the message, for example a request for
another application to load a table.
The concept behind the MType
is similar to that of a UCD \cite{ucd} 
in that a small vocabulary is sufficient to
describe the expected range of concepts required by a messaging system
within the current scope of the SAMP protocol.
Developers are free to introduce new MTypes for use within applications
without restriction; new MTypes intended to be used for Hub messaging or
other administrative purposes within the messaging system should be discussed
within the IVOA for approval as part of the SAMP standard.

\subsection{The Form of an MType}

        MType syntax is formally defined in Section \ref{sect:subscription}.
Like a UCD, an MType is made up of {\em atoms}.
These are not only meaningful to the developer, but form the central
concept of the message.  
Because the capabilities one application is searching for
are loosely coupled with the details of what another may provide,
there is not a rigorous definition of the {\em behavior} that
an MType must provoke in a receiver.  Instead, the MType defines a specific
semantic message such as ``display an image'', and it is up to the receiving
application to determine how it chooses to do the display (e.g.\ a rendered
greyscale image within an application or displaying the image in a web
browser might both be valid for the recipient and faithful to the meaning
of the message).

        The ordering of the words in an MType SHOULD normally use the
object of the message followed by the action to be performed (or the
information about that object).  For example, the use of ``{\tt image.display}''
is preferred to ``{\tt display.image}'' in order to keep the number of top-level
words (and thus message classes) like `image' small, but still allow for a
wide variety of messages to be created that can perform many useful actions
on an image.  If no existing MType exists for the required purpose,
developers can agree to the use of a new MType such as
``{\tt image.display.extnum}'' if, e.g., the ability to display a specific image
extension number warrants a new MType.

\subsection{The Description of an MType}
\label{sect:mtype-doc}

In order that senders and recipients can agree on what is meant by 
a given message, the meaning of an MType must be clearly documented.
This means that for a given MType the following information must be
available:
\begin{enumerate}
\item The MType string itself
\item A list of zero or more named parameters
\item A list of zero or more named returned values
\item A description of the meaning of the message
\end{enumerate}
For each of the named parameters, and each of the returned values,
the following information must be provided:
\begin{itemize}
\item name
\item data type ({\tt map}, {\tt list} or {\tt string}
    as described in Section \ref{sect:samp-data-types}) and if appropriate
    scalar sub-type (see Section \ref{sect:scalar-types}) 
\item meaning
\item whether it is OPTIONAL (considered REQUIRED unless stated otherwise)
\item OPTIONAL parameters MAY specify what default will be used
         if the value is not supplied
\end{itemize}
Together, this is much the same information as should be given for
documentation of a public interface method in a weakly-typed programming
language.

The parameters and return values associated with each MType 
form extensible vocabularies as explained in Section \ref{sect:vocab},
except that there is no reserved ``{\tt samp.}'' namespace.

Note that it is possible for the MType to have no returned values.
This is actually quite common if the MType does not represent a
request for data.  It is not usually necessary to define a status-like
return value (success or failure),
since this information can be conveyed as the value of the {\tt samp.status}
entry in the call response as described in Section \ref{sect:response-encode}.

\subsection{MType Vocabulary: Extensibility and Process}

The set of MTypes forms an extensible vocabulary along the lines of
Section \ref{sect:vocab}.
The relatively small set of MTypes in the ``{\tt samp.}'' namespace is
defined in Section \ref{sect:admin-mtypes} of this document,
but applications will need to use a wider range of MTypes to exchange
useful information.
Although clients are formally permitted to
define and use any MTypes outside of the reserved ``{\tt samp.}'' namespace,
for effective interoperability there must be
public agreement between application authors on this unreserved vocabulary
and its semantics.

Since addition of new MTypes is expected to be ongoing, 
MTypes from this broader vocabulary will
be documented
outside of this document to avoid the administrative
overhead and delay associated with the IVOA Recommendation Track \cite{docstd}.
At time of writing, the procedures for maintaining the list of
publicly-agreed MTypes are quite informal.
These procedures remain under review,
however the current list and details of best practice for adding to it 
are, and will remain, available in some form from the URL
\url{http://www.ivoa.net/samp/}.

\subsection{Core MTypes}
\label{sect:admin-mtypes}

This section defines those MTypes currently in the ``{\tt samp.}'' hierarchy.
These are the ``administrative''-type MTypes which are core to the
SAMP architecture or widely applicable to SAMP applications.

\newcommand{\mtypedef}[4]{
   \begin{description}
     \item[{\tt #1}:]\dbreakd
     \begin{description}
       \item[Arguments:]\dbreakd\begin{description}#2\end{description}
       \item[Return Values:]\dbreakd\begin{description}#3\end{description}
       \item[Description:]\dbreak #4
     \end{description}
   \end{description}
}

\newcommand{\mtypearg}[3]{\item[{\tt #1} ({\tt #2}) ---] #3}

\newcommand{\mtypenoargs}{\item[{\em none}]}

\subsubsection{Hub Administrative Messages}
\label{sect:hub-mtypes}

The following MTypes are for messages which SHOULD be broadcast by the hub
in response to changes in hub state.  By subscribing to these messages,
clients are able to keep track of the current set of registered applications
and of their metadata and subscriptions.
In general, non-hub clients SHOULD NOT send these messages.
\mtypedef{samp.hub.event.shutdown}
         {\mtypenoargs}
         {\mtypenoargs}
         {The hub SHOULD broadcast this message just before it exits.
          It SHOULD also send it to clients who are registered using a
          given profile if that profile is about to shut down, even if
          the hub itself will continue to operate.
          The hub SHOULD make every effort to broadcast this message even in
          case of an exit due to an error condition.}
\mtypedef{samp.hub.event.register}
         {\mtypearg{id}{string}{Public ID of newly registered client}}
         {\mtypenoargs}
         {The hub SHOULD broadcast this message every time a client 
          successfully registers.}
\mtypedef{samp.hub.event.unregister}
         {\mtypearg{id}{string}{public ID of unregistered client}}
         {\mtypenoargs}
         {The hub SHOULD broadcast this message every time a client
          unregisters.}
\mtypedef{samp.hub.event.metadata}
         {\mtypearg{id}{string}{public ID of client declaring metadata}
          \mtypearg{metadata}{map}{new metadata declared by client}}
         {\mtypenoargs}
         {The hub SHOULD broadcast this message every time a client
          declares its metadata.
          The {\tt metadata} argument is exactly as passed using the
          {\tt declareMetadata()} method.}
\mtypedef{samp.hub.event.subscriptions}
         {\mtypearg{id}{string}{public ID of subscribing client}
          \mtypearg{subscriptions}{map}{new subscriptions declared by client}}
         {\mtypenoargs}
         {The hub SHOULD broadcast this message every time a client 
          declares its subscriptions.
          The {\tt subscriptions} argument is exactly as passed using the
          {\tt declareSubscriptions()} method, and hence may contain wildcarded
          MType strings.}
\mtypedef{samp.hub.disconnect}
         {\mtypearg{reason}{string}{(OPTIONAL)
                                    Short text message indicating the reason
                                    that the disconnection is being forced}}
         {\mtypenoargs}
         {The hub SHOULD send this message to a client if the hub intends to disconnect that
          client forcibly.  This indicates that no further communication
          from that client is welcome, and any such attempts may be
          expected to fail.
          The hub may wish to disconnect clients forcibly as a result of
          some hub timeout policy or for other reasons.}

\subsubsection{Client Administrative Messages}

The following messages are generic messages defined for client use.
\mtypedef{samp.app.ping}
         {\mtypenoargs}
         {\mtypenoargs}
         {Diagnostic used to indicate whether an application is 
          currently responding.
          No ``status''-like return value is defined,
          since in general any response will indicate aliveness,
          and the normal {\tt samp.status} key in the response may be
          used to indicate any abnormal state.}
\mtypedef{samp.app.status}
         {\mtypearg{txt}{string}{Textual indication of status}}
         {\mtypenoargs}
         {General purpose message to indicate application status.}
\mtypedef{samp.app.event.shutdown}
         {\mtypenoargs}
         {\mtypenoargs}
         {Indicates that the sending application is going to shut down.
          Note that sending this message is not a substitute for unregistering
          with the hub --- registered clients about to shut down 
          SHOULD always explicitly unregister.}
\mtypedef{samp.msg.progress}
         {\mtypearg{msgid}{string}{Message ID of a previously received message}
          \mtypearg{txt}{string}{Textual indication of progress}
          \mtypearg{percent}{string}{(OPTIONAL)
                                     SAMP float value giving the approximate 
                                     percentage progress}
          \mtypearg{timeLeft}{string}{(OPTIONAL)
                                      SAMP float value giving the estimated
                                      time to completion in seconds}}
         {\mtypenoargs}
         {Reports on progress of a message previously received by the sender
          of this message.  Such progress reports MAY be sent at intervals
          between the receipt of the message and sending a reply.
          Note that the {\tt msg-id} of the earlier message must be passed to
          identify it --- the sender of the earlier message (the recipient of
          this one) will have to have retained it from the return value of the
          relevant {\tt call*()} method to match progress reports with
          requests.}

\newpage
\appendix

\section{Changes between PLASTIC and SAMP}

In order to facilitate the transition from PLASTIC to SAMP from an
application developer's point of view, we summarize in this Appendix 
the main changes. 
In some cases the reasons for these are summarized as well.

\begin{description}
\item[Language Neutrality:]
   PLASTIC contained some Java-specific ideas and details, in particular
   an API defined by a Java interface, use of Java RMI-Lite as a 
   transport protocol option, and a lockfile format
   based on java Property serialization.
   No features of SAMP are specific to, or defined with reference to, Java
   (or to any other programming language).
\item[Profiles:]
   The formal notion of a SAMP Profile replaces the choices of transport
   protocol in PLASTIC.
\item[Nomenclature:]
   Much of the terminology has changed between PLASTIC and SAMP,
   in some cases to provide better consistency with common usage in
   messaging systems.
   There is not in all cases a one-to-one correspondence betweeen PLASTIC and
   SAMP concepts, but a partial translation table is as follows:
   \begin{center}
   \begin{tabular}{ll}
   PLASTIC                &  SAMP  \\
   \hline
   message                &  MType         \\
   support a message      &  subscribe to an MType  \\
   registered application &  client \\
   synchronous request    &  synchronous call/response  \\
   asynchronous request   &  notification 
   \end{tabular}
   \end{center}
\item[MTypes:]
   In PLASTIC message semantics were defined using opaque URIs such as
   {\tt ivo://votech.org/hub/event/HubStopping}.  
   SAMP replaces these with a vocabulary of structured MTypes
   such as {\tt samp.hub.event.shutdown}.
\item[Asynchrony:]
   Responses from messages in PLASTIC were returned synchronously, using
   blocking methods at both sender and recipient ends.  As well as 
   inhibiting flexibility, this risked timeouts for long processing times
   at the discretion of the underlying transport.
   The basic model in SAMP relies on asynchronous responses, though 
   a synchronous fa\c{c}ade hub method is also provided for 
   convenience of the sender.
   Client toolkits may also wish to provide client-side synchronous 
   fa\c{c}ades based on fully asynchronous messaging.
\item[Registration:]
   In PLASTIC clients registered with a single call which acquired a
   hub connection and declared callback information, message subscriptions, 
   and some metadata.
   In SAMP, these four operations have been decomposed into separate calls.
   As well as being tidier, this offers benefits such as meaning that
   the subscriptions and metadata can be updated during the lifetime of
   the connection.
\item[Client Metadata:]
   PLASTIC stored some application metadata (Name) in the hub and provided
   acess to others (Description, Icon URL, \ldots) using custom messages.
   SAMP stores it all in the hub providing better extensibility and consistency
   as well as improving metadata provision for non-callable applications
   and somewhat reducing traffic and burden on applications.
\item[Named Parameters:]
   The parameters for PLASTIC messages were identified by sequence
   (forming a list), while
   the parameters for SAMP MTypes are identified by name (forming a map).
   As well as improving documentability, this makes it much more convenient
   to allow for optional parameters or to introduce new ones.
   The same arrangement applies to return values.
\item[Recipient Targetting:]
   PLASTIC featured methods for sending messages to all or to an explicit list 
   of recipients.  In practice the list variants were rarely used except to
   send to a single recipient.
   SAMP has methods for sending to all or to a single recipient.
\item[Typing:]
   Data types in PLASTIC were based partly on Java and partly on XML-RPC types.
   There was not a one-to-one correspondence between types in the Java-RMI
   transport and the XML-RPC one, which encouraged confusion.
   Parameter types included integer, floating point and boolean as well as
   string, which proved problematic to use correctly from some 
   weakly-typed languages.
   SAMP uses a more restricted set of types 
   (namely string, list and map) at the protocol level,
   along with some auxiliary rules for encoding numbers and booleans as strings.
\item[Lockfile:]
   The lockfile in SAMP's standard profile is named {\tt .samp},
   its format is defined explicitly rather than with reference
   to Java documentation, and there is better provision for its 
   location in a language-independent way on MS Windows systems.
   In many cases however, the same lockfile location/parsing code 
   will work for both SAMP and PLASTIC except for the different filenames
   (``.samp'' vs.\ ``.plastic'').
\item[Public/Private ID:]
   In PLASTIC a single, public ID was used to label and identify 
   applications during communications directed to the hub
   or to other applications.
   This meant that applications could easily, if they wished, 
   impersonate other applications.  The practice in SAMP is to
   use different IDs for public labelling and private identification,
   which means that such ``spoofing'' is no longer a danger.
\item[Errors:]
   SAMP has provision to return more structured error information 
   than PLASTIC did.
\item[Extensibility:]
   Although PLASTIC was in some ways extensible, SAMP provides more hooks
   for future extension, in particular by pervasive use of the 
   {\em extensible vocabulary\/} pattern.
\end{description}

\section{Change History}
\label{sect:changes}

Changes to SAMP between Working Draft version 1.0 (2008-06-25) and
Recommendation version 1.11 (2009-04-21):
\begin{itemize}
\item Return values of {\tt callAll} and {\tt notifyAll} operations changed;
      they now return information about clients receiving the messages
      (Section \ref{sect:hubOps}).
\item Characters allowed in {\tt string} type restricted to avoid
      problems transmitting over XML;
      was 0x01--0x7f, now 0x09, 0x0a, 0x0d, 0x20--0x7f
      (Section \ref{sect:samp-data-types}).
\item New hub administrative message {\tt samp.hub.disconnect}
      (Section \ref{sect:hub-mtypes}).
\item Empty placeholder appendix on SAMP/PLASTIC interoperability removed.
\item Wording clarified and made more explicit in a few places.
\item Typos fixed, including incorrect BNF in Section \ref{sect:subscription}.
\item Author list re-ordered.
\item Editorial changes and clarifications following RFC period.
\item MType Vocabulary section now directs readers to
      {\tt http://www.ivoa.net/samp/} to find current MType list and process.
\end{itemize}

Changes to SAMP between Recommendation version 1.11 (2009-04-21) and
version 1.2 (2010-12-16):
\begin{itemize}
\item Use of new SAMP\_HUB environment variable lockfile location option
      documented in section \ref{sect:lockfile}.
\item Added Non-Technical Preamble section \ref{sect:nonTechPreamble}
      as per agreement for all new/revised IVOA documents.
\end{itemize}

Changes to SAMP between Recommendation version 1.2 (2010-12-16) and
version 1.3 (2012-04-11):
\begin{itemize}
\item Add a new Section \ref{sect:webprofile} on the Web Profile.
      Minor changes in the rest of the document noting the existence
      of this new Profile.
\item Add a new Section \ref{sect:security} discussing security issues
      in general, with reference to their particular consideration for
      both Standard and Web Profiles.  The discussion of Standard Profile
      security is moved to its own new Section \ref{sect:std-security}.
\item MType syntax declaration in Section \ref{sect:subscription} now permits
      upper-case letters (for consistency with actual usage).
\item Sections \ref{sect:response-encode} and \ref{sect:faults} now note
      that the hub is permitted to generate and forward an error response
      on behalf of a client under some circumstances.
      The {\tt samp.code=samp.noresponse} code is reserved for
      this purpose.
\item Section \ref{sect:vocab} now reserves a namespace ``{\tt x-samp}''
      for keys in an extensible vocabulary which are proposed
      for possible future introduction into this standard.
\item A comment has been added to Section \ref{sect:samp-data-types}
      concerning recommended protocols for use with URLs in messages.
\end{itemize}


\end{document}